\let\ifarxiv=\iftrue     
\pdfoutput=1

\ifarxiv

\documentclass[12pt,a4paper]{article}
\usepackage[a4paper,text={450pt,650pt},centering]{geometry}

\fi

\ifarxiv\else

\documentclass[11pt,a4paper]{article}
\usepackage{mathptmx}
\usepackage[a4paper,text={130mm,198mm}]{geometry}

\fi

\usepackage{amsmath,amssymb}
\usepackage[bookmarks=true,hyperfigures=true]{hyperref}
\usepackage{graphicx}
\usepackage[nosort]{cite}
\usepackage[bulletsep]{collref}

\usepackage{amsfonts}

\let\oldbfseries=\bfseries
\let\oldmdseries=\mdseries
\let\oldnormalfont=\normalfont
\renewcommand{\bfseries}{\oldbfseries\boldmath}
\renewcommand{\mdseries}{\oldmdseries\unboldmath}
\renewcommand{\normalfont}{\oldnormalfont\unboldmath}

\allowdisplaybreaks[3]

\numberwithin{equation}{section}

\usepackage[font=small,labelfont=bf,width=0.85\textwidth]{caption}

\providecommand{\hypersetup}[1]{}
\providecommand{\texorpdfstring}[2]{#1}

\hypersetup{plainpages=false}
\hypersetup{pdfpagemode=UseNone}
\hypersetup{bookmarksnumbered=true}
\hypersetup{pdfstartview=FitH}
\hypersetup{colorlinks=false}
\hypersetup{citebordercolor={.5 1 .5}}
\hypersetup{urlbordercolor={.5 1 1}}
\hypersetup{linkbordercolor={1 .7 .7}}

\providecommand{\href}[2]{#2}
\providecommand{\arxivlink}[1]{\href{http://arxiv.org/abs/#1}{arxiv:#1}}

\newcommand{\be}{\begin{equation}}
\newcommand{\ee}{\end{equation}}

\newcommand{\half}{\frac12}
\newcommand{\ra}{\rightarrow}

\newcommand{\Urm}{\mathrm{U}}
\newcommand{\SU}{\mathrm{SU}}
\newcommand{\SO}{\mathrm{SO}}

\newcommand{\SL}{\mathrm{SL}}

\newcommand{\Tr}{\mathrm{Tr}}
\newcommand{\Wcal}{\mathcal{W}}
\newcommand{\mparagraph}[1]{\paragraph{#1}\mbox{}\vspace{.04cm}}

\newcommand{\Zset}{\mathbb{Z}}
\newcommand{\Ncal}{\mathcal{N}}

\newcommand{\AdS}{\mathrm{AdS}}
\newcommand{\Srm}{\mathrm{S}}
\newcommand{\Cset}{{\,\,{{{^{_{\pmb{\mid}}}}\kern-.47em{\mathrm C}}}}}

\newcommand{\diff}{\mathrm{d}}

\newcommand{\comment}[1]{}

\begin{document}


\thispagestyle{empty}
\phantomsection
\addcontentsline{toc}{section}{Title}

\begin{flushright}\footnotesize%
\texttt{\arxivlink{1012.3998}}\\
overview article: \texttt{\arxivlink{1012.3982}}%
\vspace{1em}%
\end{flushright}

\begingroup\parindent0pt
\begingroup\bfseries\ifarxiv\Large\else\LARGE\fi
\hypersetup{pdftitle={Review of AdS/CFT Integrability, Chapter IV.2: Deformations, Orbifolds and Open Boundaries}}%
Review of AdS/CFT Integrability, Chapter IV.2:\\
 Deformations, Orbifolds and Open Boundaries
\par\endgroup
\vspace{1.5em}
\begingroup\ifarxiv\scshape\else\large\fi%
\hypersetup{pdfauthor={Konstantinos Zoubos}}%
Konstantinos Zoubos
\par\endgroup
\vspace{1em}
\begingroup\itshape
Niels Bohr Institute, Blegdamsvej 17, DK-2100 Copenhagen \O, Denmark
\par\endgroup
\vspace{1em}
\begingroup\ttfamily
kzoubos@nbi.dk
\par\endgroup
\vspace{1.0em}
\endgroup

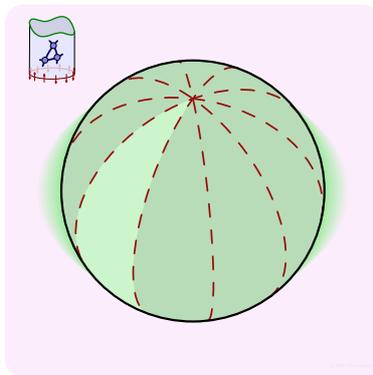
\begin{center}
\includegraphics[width=5cm]{TitleIV2.mps}
\vspace{1.0em}
\end{center}

\paragraph{Abstract:}
We review the role of integrability in the planar spectral problem of four-dimensional superconformal
gauge theories besides $\Ncal=4$ SYM. The cases considered include the Leigh--Strassler marginal 
deformations of $\Ncal=4$ SYM, quiver theories which arise as orbifolds of $\AdS_5\times\Srm^5$ on the 
dual gravity side, as well as various theories involving open spin chains.  

\ifarxiv\else
\paragraph{Mathematics Subject Classification (2010):} 
81R50, 81T30, 81T60, 82B23
\fi
\hypersetup{pdfsubject={MSC (2010): 81R50, 81T30, 81T60, 82B23}}%

\ifarxiv\else
\paragraph{Keywords:} 
Integrability, AdS/CFT, Bethe Ansatz, Marginal Deformations, Orbifolds, Open Spin Chains
\fi
\hypersetup{pdfkeywords={Integrability, AdS/CFT, Bethe Ansatz, Marginal Deformations, Orbifolds, Open Spin Chains}}%

\newpage



\section{Introduction}

The fascinating integrable structures of the $\Ncal=4$ SYM theory, reviewed in
other contributions to this collection, highlight the unique position that
this theory occupies among quantum field theories in four dimensions. Planar integrability 
is just the latest addition to a long list of remarkable properties, such as exact (perturbative 
and non-perturbative) conformal invariance, Montonen--Olive S-duality, as well as the celebrated 
AdS/CFT correspondence, stating its equivalence to IIB string theory on the $\AdS_5\times \Srm^5$ background.

The price to pay for these unique features is that the theory is highly unrealistic,
and arguably very far removed from QCD, the theory of the strong interactions. 
It is thus natural to wonder whether the recent great advances in the understanding 
of $\Ncal=4$ SYM are of any use when studying less supersymmetric theories. In the 
specific context of AdS/CFT integrability, 
one can ask whether there exist other four-dimensional field theories with similar integrability 
structures, where the techniques developed in the $\Ncal=4$ SYM context can also be
applied. 

In this short review we will attempt to provide a guide to the current state of affairs
regarding AdS/CFT integrability in less supersymmetric situations. We will
restrict ourselves to the very special class of four-dimensional supersymmetric field theories with
 similar finiteness properties to $\Ncal=4$ SYM, which are therefore also superconformal.\footnote{We 
will thus not touch the topic of integrability
in QCD, which is covered in \cite{chapQCD} in this collection. Neither will we discuss 
integrability in the 3-dimensional ABJM theory, referring instead to \cite{chapN6}.} 
We will see that, despite many similarities to the $\Ncal=4$ SYM case, there also appear significant
differences in the way integrability is manifested.  
Therefore, although there still is quite a long way to go from these theories to QCD, 
their study is worthwhile and can be expected to provide a useful stepping-stone towards 
unraveling the implications of integrability in more realistic field theories.

\section{The Marginal Deformations of \texorpdfstring{$\Ncal=4$}{N=4} SYM}

For any conformal field theory, it is interesting to study its space of \emph{exactly}  
\emph{marginal deformations}, all the ways to deform the theory preserving quantum conformal
invariance. It has been known since the early eighties that $\Ncal=4$ SYM admits $\Ncal=1$ 
supersymmetric marginal deformations, with a non-perturbative proof given by Leigh and Strassler in
1995 \cite{Leigh:1995ep} (where references to the earlier literature can also be found).

In $\Ncal=1$ superspace language, the Leigh--Strassler deformations can be obtained purely 
by deforming the superpotential of the $\Ncal=4$ SYM theory. 
The relevant part of the $\Ncal=4$ SYM lagrangian is (with $g$ being the gauge coupling)
\be \label{N=4lagrangian}
\mathcal{L}_{\mathrm{sup}}=\int\mathrm{d}^2\theta\; \Wcal_{\Ncal=4}\;,\quad\text{where} \quad \Wcal_{\Ncal=4}=
g {\mathrm {Tr}}(X[Y,Z])\;.
\ee
Here $X,Y$ and $Z$ are the three adjoint chiral superfields of the $\Ncal=4$ theory. 
It is not hard to see that $\Wcal_{\Ncal=4}$ possesses an $\SU(3)\times \Urm(1)_R$ global invariance,
the maximal part of the $\SU(4)$ R-symmetry of the theory which can be made explicit
in $\Ncal=1$ superspace. 
Now consider the following more general $\Ncal=1$ superpotential:
\be \label{LSlagrangian}
\Wcal_{LS}=
\kappa{\mathrm {Tr}}\left(X[Y,Z]_{q}+ \frac h3\left(X^3\!+\!Y^3\!+\!Z^3
\right)\right)
\ee 
where $\kappa, q$ and $h$ are a priori complex parameters and the $q$-commutator is defined as $[X,Y]_q=XY-qYX$. 
The $\Ncal=4$ SYM theory can be 
recovered by the choice $(\kappa,q,h)=(g,1,0)$. Generically, the only continuous symmetry of $\Wcal_{LS}$ is
the $\Urm(1)_R$ which is always present in an $\Ncal=1$ superconformal theory.
When $h=0$, it is standard to express $q=\exp(2\pi i\beta)$ and call
this case the $\beta$-$deformation$.\footnote{There exist several other conventions in the 
literature, related by relabellings of $\beta$ and $\kappa$.} 
Here, apart from $\Urm(1)_R$, $\Wcal_{LS}$ has an extra $\Urm(1)\times \Urm(1)$ symmetry 
acting by phase rotations on the scalars.
The $\beta$-deformation with $\beta$ real (i.e. $q$ a phase) is variously known as the 
real-$\beta$ or the $\gamma$-deformation. 

There are several more marginal terms one could add to 
the superpotential, however (\ref{LSlagrangian}) is special in that it preserves an 
important set of discrete symmetries:
\be\begin{split}
\text{(a)}\;\;& X\!\rightarrow Y\;,\;Y\!\rightarrow Z\;,\;Z\!\rightarrow X,\;\;\\
\text{(b)}\;\;& X\!\rightarrow X\;,\;Y\!\rightarrow \omega Y\;,\;Z\!\rightarrow\omega^2 Z
\end{split}
\ee
with $\omega$ a third root of unity. The first of these symmetries is particularly crucial, because 
it ensures that all scalar anomalous dimensions are equal. This observation
allowed Leigh and Strassler to argue that finiteness imposes a \emph{single} 
complex constraint on the four couplings $(g,\kappa, q, h)$,
implying the existence of a three-dimensional parameter space of finite gauge theories. 
On this space, the superpotential (\ref{LSlagrangian}) is thus \emph{exactly} marginal. The finiteness
constraint can be calculated at low loop orders, but its exact form
is unknown, and its determination, even in the planar limit, would be a major step in our
understanding of superconformal gauge theory. 
Here we give it at one loop (see e.g. \cite{Aharony:2002tp} for a derivation):
\be \label{1loop}
2g^2=\kappa\bar{\kappa}\left[\frac{2}{N^2}(1+q)(1+\bar{q})+\left(1-\frac{4}{N^2}\right)(1+q\bar{q}+h\bar{h})\right]\;.
\ee
Note that the constraint simplifies considerably in the planar ($N\ra \infty)$ limit, and that for the 
real $\beta$-deformation it reduces to $g^2=\kappa\bar{\kappa}$, precisely the same as that for 
$\Ncal=4$ SYM. It has been shown \cite{Mauri:2005pa} that in this real-$\beta$ case the one-loop constraint 
is not modified at any higher order in the perturbative expansion. This
is a first indication that, in the planar limit, the theory will share many of the properties 
of $\Ncal=4$ SYM, including, as we will see, integrability.

\subsection{The gravity dual of the \texorpdfstring{$\beta$}{beta}-deformation}

If the $\Ncal=4$ SYM theory admits exactly marginal deformations, the same must be true for its dual gravity
background. Since the deformations preserve the conformal group, the $\AdS_5$ factor of the geometry will
not be affected, but we expect the $\Srm^5$ part to be deformed, reflecting the reduction of
the R-symmetry group to a subgroup of $\SU(4)_R\simeq \SO(6)$. 
 For the $\beta$-deformation, the metric of this deformed $\Srm^5$ was found by Lunin and Maldacena 
in 2005 \cite{Lunin:2005jy}. Focusing first on the \emph{real}-$\beta$ deformation, these authors showed that 
it can be obtained from $S^5$ by a sequence of T-duality, angle shift and T-duality, called a $TsT$ transformation.
To make this a bit more explicit, let us start with the 5-sphere embedded in ${\mathbb R}^6$ as $\bar X X+\bar Y Y+\bar Z Z=1$,
and parametrise
\be
X=\cos\gamma e^{i\varphi_1}\;,\quad Y=\sin\gamma \cos\psi e^{i\varphi_2}\;,\quad 
Z=\sin\gamma \sin\psi e^{i\varphi_3}
\ee
to obtain the five-sphere metric in terms of angle coordinates
\be \label{5sphere}
\diff s^2=\diff \gamma^2+\cos^2\gamma \diff \varphi_1^2+\sin^2\gamma
\left(\diff \psi^2+\cos^2\psi\diff\varphi_2^2+\sin^2\psi\diff \varphi_3^2\right).
\ee
There are three explicit $\Urm(1)$ isometries related to the angles $\varphi_i$, with the diagonal shift
$\varphi_i\ra \varphi_i+a$ corresponding to the $\Urm(1)_R$ which is required by $\Ncal=1$ superconformal 
invariance. The $TsT$ procedure starts by T-dualising along the other two isometry
directions, then shifting the dual angles as $\tilde\varphi_2\rightarrow\tilde\varphi_2+\beta\tilde\varphi_3$, 
and finally T-dualising back. This breaks the $\SO(6)$ implicit in (\ref{5sphere}) and results in a geometry 
preserving just a $\Urm(1)^3$ isometry group, the right amount of symmetry for the dual to the $\beta$-deformation. 
We refer to 
\cite{Lunin:2005jy,Frolov:2005dj} for more details and for the explicit IIB 
solution.\footnote{It should also be noted that for $\beta=1/k$ (i.e. $q$ being a $k$-th root of unity) 
the dual background is actually an $\AdS_5\times \Srm^5/\Zset_k\times \Zset_k$ orbifold \cite{Berenstein:2000ux}.}
Starting from the real-$\beta$ background, a sequence of S-dualities leads to the dual of the 
complex-$\beta$ deformation \cite{Lunin:2005jy}. However, the geometry dual to
the most general deformation (with $h\neq 0$) is still unknown.

\subsection{The real-\texorpdfstring{$\beta$}{beta} deformation and integrability}

In this section we focus on the real-$\beta$ deformation, which has received the 
most attention in the literature. The integrability properties of this theory were first 
investigated in \cite{Roiban:2003dw}, where it was shown that the one-loop planar 
dilatation generator in the two-scalar $\SU(2)_\beta$ sector 
corresponds to the hamiltonian of the integrable XXX $\SU(2)_\beta$ spin chain. 
This was extended to the $\SU(3)_\beta$ sector in \cite{Berenstein:2004ys}.
 In the latter work it was also noted that a suitable site-dependent transformation 
can map the hamiltonian of the deformed theory to that of the undeformed
one (i.e. $\Ncal=4$ SYM) but with \emph{twisted} boundary conditions. 
Building on \cite{Lunin:2005jy}, where a simple star-product was introduced to keep track
of the additional phases appearing in the real-$\beta$ theory compared to the undeformed case, the
work \cite{Beisert:2005if} showed that given an undeformed $R$-matrix satisfying the
Yang--Baxter equation, the twisted one will do so as well.\footnote{The effect of the twist 
on other algebraic structures of the theory, such as the Yangian (reviewed in \cite{chapYang}), 
was considered in \cite{Ihry:2008gm}.}  

The conclusion is that the real-$\beta$ deformation is just as integrable as $\Ncal=4$ SYM.
It should thus be possible to find a Bethe ansatz encoding the spectrum of the theory. This
can indeed be done by introducing appropriate phases (``twisting'') in the $\Ncal=4$ SYM Bethe ansatz.  
In the $\SU(2)_\beta$ sector, this was performed at one loop in \cite{Berenstein:2004ys},
at higher loops in \cite{Frolov:2005ty}, while the higher-loop twist for all sectors was obtained 
in \cite{Beisert:2005if}. For simplicity, here we display just the one-loop, $\SU(2)_\beta$ 
sector case:
\be \label{LSBetheansatz}
e^{-2\pi i \beta L}\left(\frac{u_k+i/2}{u_k-i/2}\right)^L
=\prod_{j=1,j\neq k}^{M}\frac{u_k-u_j+i}{u_k-u_j-i}\;,\qquad 
\prod_{k=1}^{M}\frac{u_k+i/2}{u_k-i/2}=e^{2\pi i \beta M}
\ee
where the second equation is the cyclicity condition. Very recently, \cite{Ahn:2010ws} 
provided a deeper understanding of the all-loop-twisted Bethe equations by deriving them 
from a suitable \emph{Drinfeld-twisted} S-matrix combined with a twist of the boundary conditions.

\mparagraph{Integrability and the LM background}
 
Integrability of the IIB Green--Schwarz sigma model in the real-$\beta$ deformed case was demonstrated in 
\cite{Frolov:2005dj} by explicit construction of a Lax pair for the LM 
background. A Lax pair was also constructed for the pure-spinor sigma-model in \cite{Grassi:2006tj}.
Therefore, just as in the undeformed case (reviewed in \cite{chapSpinning,chapQString}) one can attempt to 
compare gauge theory results with strong coupling ones by considering semiclassical strings moving on the 
LM background. This was done in \cite{Chen:2005sba,Chen:2006bh}, with the construction of several 
semiclassical string solutions, which were matched to specific configurations of roots of the twisted Bethe 
ansatz. Their energies precisely agree with the gauge theory anomalous dimensions. 

Giant magnons \cite{Hofman:2006xt} on the LM background were constructed in 
\cite{Chu:2006ae} and \cite{Bobev:2005cz}, with the latter considering multispin configurations, while 
\cite{Bobev:2007bm} considered more general rigid string solutions on the $\Srm^3_\gamma$ subspace, 
with the giant magnons and spiky strings as special cases. 
The first finite-size correction to the giant magnon energy was computed in
\cite{Bykov:2008bj} and takes the following form:\footnote{Recently, this result was extended to the case of
dyonic, or two-spin, giant magnons \cite{Ahn:2010da}.}
\be
E-J=2g\sin\frac p2\left(1-\frac{4}{e^2}\sin^2\frac p 2 \cos\left[\frac{2\pi(n-\beta J)}
{2^{3/2}\cos^3\frac p 4}\right] e^{-\frac{J}{g\sin p/2}}+\cdots \right)
\ee
where $n$ is the unique integer for which $|n-\beta J|\leq\half$. 
This expression exhibits the expected exponential falloff, but the momentum dependence is
highly unusual, and indeed reproducing  it from the L\"uscher correction techniques discussed 
in \cite{chapLuescher} is still an open problem.

\mparagraph{Wrapping corrections}

In order to calculate wrapping corrections to the spectrum (due to interactions whose span is greater than the length of the spin
chain), one needs to go beyond the asymptotic Bethe ansatz. It turns out that the techniques developed for
$\Ncal=4$ SYM (reviewed in \cite{chapLuescher,chapTBA,chapTrans} in this collection) can be applied with relative
ease to the $\beta$-deformed theory. In particular, it was argued in \cite{Gromov:2010dy} that the $\beta$-deformation
is described by the \emph{same} Y-system as $\Ncal=4$ SYM. The $\beta$ parameter arises by appropriately modifying
the asymptotic (large $L$) solution, exploiting the freedom to twist it by certain complex numbers. The authors of
\cite{Gromov:2010dy} showed that this procedure correctly reproduces the higher-loop asymptotic Bethe ansatz of 
\cite{Beisert:2005if} (for all sectors, and more general twists) and derived generalised L\"uscher 
formulae for generic operators in the $\beta$-deformed theory.

Turning to results for specific operators, an interesting feature of the $\beta$-deformed theory 
compared to $\Ncal=4$ SYM (first noted in \cite{Freedman:2005cg}) is that one-impurity operators
\be
O_{1,L}=\Tr\phi Z^{L-1}\;,\qquad \phi\in\{X,Y\}
\ee
are not protected by 
supersymmetry and thus acquire anomalous dimensions. Because of this, the real-$\beta$ theory provides an excellent setting
for the perturbative study of wrapping effects for short operators (reviewed in \cite{chapHigher} and also in
\cite{Fiamberti:2010fw}): Apart from the calculations 
being simpler (compared to two-impurity cases like the Konishi operator), it also allows for 
a clean separation of the effects of wrapping from those
due to the dressing factor, since the latter does not contribute at all for these states. Wrapping effects
for such states, at critical wrapping (where the loop level equals the length of the operator) have been computed 
up to 11 loops in \cite{Fiamberti:2008sm,Fiamberti:2008sn} (who also provided a recursive formula for
calculating them at higher loop orders), and have recently been reproduced
in \cite{Gromov:2010dy} via the twisted solution to the Y-system and in \cite{Arutyunov:2010gu} using generalised
L\"uscher formulae.\footnote{Note that no TBA equations (see \cite{chapTBA})  have yet been constructed for 
the $\beta$-deformed theory.}

A very special case arises when $\beta=\half$ and one considers even length operators. 
Then the (higher-loop version of the) Bethe ansatz (\ref{LSBetheansatz}) 
becomes the same as that for $\Ncal=4$ SYM, apart from a sign in the cyclicity constraint. In this case,
a closed (instead of iterative) form for the critical wrapping
correction at any $L$ was found in  \cite{Beccaria:2009hg}. Also working at $\beta=\half$,
and using the L\"uscher techniques reviewed in \cite{chapLuescher}, the work \cite{Bajnok:2009vm} 
calculated the wrapping corrections to the single-impurity operator with $L=4$ up to five
loops, i.e. the first two nontrivial orders:\footnote{Here $\Delta_w$ denotes the wrapping contribution
to the anomalous dimension, i.e. the difference of the exact result from the asymptotic one.}
\be\begin{split}
\Delta_w^{\text{4-loop}}=&128(4\zeta(3)-5\zeta(5)),\\
\Delta_w^{\text{5-loop}}=&-128(12\zeta(3)^2+32\zeta(3)+40\zeta(5)-105\zeta(7))\;.
\end{split}
\ee
The four-loop result agrees with the perturbative calculations in \cite{Fiamberti:2008sm}, while
at the time of writing there does not exist a perturbative result for the five-loop 
(subleading wrapping) correction. In \cite{Gunnesson:2009nn}, the wrapping corrections at 
$\beta=\half$ were used to argue for the equivalence (suggested by (\ref{LSBetheansatz}) for the asymptotic
spectrum) of the full (non-asymptotic) spectra of the $\beta$-deformed theory at $\beta$ and $\beta+1/L$, with
the recent leading-finite-size results of \cite{Arutyunov:2010gu} in complete agreement with this. 

Moving on to the two $L=4$ two-impurity operators ($\Tr(XYXY)$ and $\Tr(XXYY)$), their anomalous dimensions were found to 
four-loop order through explicit calculation 
in \cite{Fiamberti:2008sm,Fiamberti:2008sn}.\footnote{Note that in $\Ncal=4$ SYM one linear combination of these operators
is BPS, while the other is a descendant of the Konishi operator. However, in the 
deformed theory there is no BPS combination.} They were also computed and matched (for arbitrary $\beta$) using
L\"uscher methods in \cite{Ahn:2010yv} as well as through the Y-system in \cite{Gromov:2010dy}. Essentially
the same calculation (starting from a slightly different perspective) was performed in \cite{Arutyunov:2010gu}.

Finally, there exists at the moment a prediction \cite{Arutyunov:2010gu}, coming from L\"uscher methods, 
for the leading finite-size correction to the energy for one- and two- impurity $sl(2)$-sector operators, 
which has yet to be checked by explicit perturbative calculations.\footnote{See also \cite{Beccaria:2010kd,deLeeuw:2010ed} for
more recent results on wrapping for twist operators in the $\beta$-deformed theory.}

\mparagraph{Amplitudes}

As reviewed in \cite{chapAmp}, one manifestation of integrability in the $\Ncal=4$ SYM context
is the appearance of iterative structures (which go by the name of the \emph{BDS conjecture}) expressing multiloop 
amplitudes in terms of one-loop ones. One might therefore expect that amplitudes
in the real-$\beta$ theory satisfy such relations as well. It has indeed been shown \cite{Khoze:2005nd}
that all (MHV and non-MHV) planar amplitudes in the real-$\beta$ theory are proportional to the 
corresponding $\Ncal=4$ SYM ones, differing only in phases affecting the tree-level part of the 
amplitude. Thus the BDS conjecture for $\Ncal=4$ SYM extends straightforwardly to the real-$\beta$ 
deformation. At strong coupling (where the tree-level part is not visible), gluon amplitudes 
in the real-$\beta$ theory have been shown to equal those for $\Ncal=4$ SYM \cite{Oz:2007qr}.

\subsection{Integrability beyond the real-\texorpdfstring{$\beta$}{beta} deformation?}

 In the above we focused on a very special subset of the marginal deformations, those where $h=0$, while
$q$ is just a phase. One can ask whether there exist other integrable values of the parameters $(q,h)$. Keeping
$h=0$ but passing to complex $\beta$, the hamiltonian in the two-holomorphic-scalar 
 sector is that of the $\SU(2)_q$ XXZ model and is thus integrable \cite{Roiban:2003dw}. However, 
this generically ceases to be the case beyond this
simple sector \cite{Berenstein:2004ys}: Contrary to initial expectations, the one-loop hamiltonian
in the \emph{full} scalar sector is not that of the integrable $\SO(6)_q$ XXZ spin chain,
but of a type not matching any known integrable hamiltonians. It
was also shown in \cite{Berenstein:2004ys} that, unlike the real-$\beta$ case, it is not possible to 
transfer the deformation to the boundary conditions
by site-dependent redefinitions.\footnote{Note, furthermore, that the star-product techniques of \cite{Lunin:2005jy} 
do not apply beyond real $\beta$, their naive extension giving rise to a non-associative product.} 
The conclusion was that the one-loop hamiltonian for the \emph{generic} LS deformation is
not integrable.\footnote{The question of whether higher-loop integrability
persists in the (all-loop closed) $\SU(2)_q$ sector remains open, with some progress towards constructing
the required higher charges reported in \cite{Mansson:2007jg}.}

Nevertheless, as demonstrated in \cite{Bundzik:2005zg}, there \emph{do} exist certain special choices of 
parameters for which the one-loop hamiltonian is integrable:
\be \label{Integrablecases}
(q,h)=\left\{(0,1/\bar{h})\;,\;\left((1+\rho)\, e^{\frac{2\pi i m}{3}},\rho\, e^{\frac{2\pi i n}{3}}\right)\;,
\;\left(-e^{\frac{2\pi i m}{3}},e^{\frac{2\pi i n}{3}}\right)\right\}\;.
\ee
Some of these choices were also discovered via the study of amplitudes in \cite{Bork:2007bj}:
They correspond to special cases where the 1-loop planar finiteness condition (\ref{1loop}) does not receive 
corrections at higher loops, similarly to the real-$\beta$ deformation. 

 In \cite{Mansson:2008xv}, a unifying framework for all these integrable cases was proposed: Their 
corresponding one-loop hamiltonians can be related to the real-$\beta$ case by \emph{Hopf twists}.
These are a way of modifying the underlying $R$-matrix, leaving the integrability properties unaffected. 
Since (as shown in \cite{Beisert:2005if}) the real-$\beta$ hamiltonian is itself related to the 
undeformed hamiltonian by such a twist, all these integrable cases can be seen to be nothing 
but Hopf-twisted $\Ncal=4$ SYM.  

Another special (one-loop) integrable sector beyond real $\beta$ was found in \cite{Mansson:2007sh}: 
It is an $\SU(3)$ sector composed of two holomorphic and one antiholomorphic scalar, 
for instance $\{X,Y,\bar{Z}\}$. The hamiltonian in this sector actually turns out to be XXZ 
$\SU(3)_q$, the standard (integrable) $q$-deformation of $\SU(3)$. However, this
sector is not closed beyond one loop, complicating the discussion of higher-loop integrability. 

 Apart from these special cases, the deformed hamiltonian is not integrable.  
An intuitive explanation for this \cite{Frolov:2005ty} is that the construction of the dual 
gravity background for the complex $\beta$ deformation involves a sequence of S-duality
transformations on the LM background \cite{Lunin:2005jy}. The strong-weak nature of S-duality means 
that the intermediate stages involve interacting strings, which (as reviewed in 
\cite{chapObserv}) are unlikely to preserve integrability. 

A more direct argument for this lack of integrability was recently given in \cite{Mansson:2008xv}, 
who noticed that there exists a Hopf algebraic deformation of
the global $\SU(3)$ R-symmetry group of the $\Ncal=4$ theory under which the full LS superpotential 
(\ref{LSlagrangian}) is invariant. However, this symmetry, defined through a 
suitable $R$-matrix depending on the deformation parameters $q$ and $h$, is not a ``standard'' quantum-group deformation
of $\SU(3)$. In particular, apart from the special cases discussed above, the $(q,h)$--$R$-matrix does \emph{not} 
respect the Yang--Baxter equation, and consequently the corresponding Hopf algebra is not quasitriangular. 
Thus the construction (reviewed in \cite{chapABA}) of the transfer matrices and eventually of the 
integrable S-matrix of the theory would not be expected to go through.

\subsubsection{More general TsT transformations}

A different way of generalising the Lunin--Maldacena solution is by performing TsT transformations along
all three available $\Urm(1)$'s within the $\Srm^5$ \cite{Frolov:2005dj}. Since one of these corresponds 
to the R--symmetry, this procedure will completely break the superconformal symmetry. However it can be shown 
that these \emph{$\gamma_i$-deformations} preserve integrability: The Lax pair construction goes through in this
case as well \cite{Frolov:2005dj} and in \cite{Alday:2005ww} it was argued that the Green--Schwarz action 
on TsT-deformed
backgrounds is the same as the undeformed one, but with twisted boundary conditions. In \cite{Frolov:2005iq}, string 
energies were shown to match  anomalous dimensions coming from the corresponding three-phase deformed spin chain. 
In addition, \cite{Prinsloo:2005dq} showed that the action  for three-spin strings in the ``fast string'' limit
admits
a Lax pair and thus that string motion is integrable. The integrability properties of the $\gamma_i$ theories
are thus very similar to the real-$\beta$ case.\footnote{As was the case for the $\beta$ deformation, it is possible to
generalise the $\gamma_i$-deformations to complex values of $\gamma_i$ while preserving integrability, but only 
for very special
values, similar to (\ref{Integrablecases}) \cite{Freyhult:2005ws}.}

One can also perform integrability-preserving $TsT$ transformations along one $AdS_5$ and one $\Srm^5$
direction, leading to dipole-type deformations in the gauge theory \cite{Bobev:2005ng},
as well as purely along the $AdS_5$ directions, leading to a noncommutative deformation
on the gauge theory side \cite{McLoughlin:2006cg} (see \cite{Swanson:2007dh} for a
review of the latter case).

\subsubsection{Non-field theory deformations}

As was first noted in \cite{Berenstein:2004ys}, there exist integrable deformations of the
algebraic structures at the $\Ncal=4$ SYM point which do not have a good field theory
interpretation, in the sense of arising as the one-loop hamiltonian of a deformed field
theory. A large class of such deformations was presented in \cite{Beisert:2005if}. More
recently, $q$-deformations of the $\mathrm{psu}(2|2)\ltimes {\mathbb R}^3$ algebra were 
considered in \cite{Beisert:2008tw}. 
The role of such deformations in the AdS/CFT correspondence is not well understood, but their 
further study can be expected to provide a deeper understanding of the $\Ncal=4$ integrable structures by 
embedding them in a larger framework.\footnote{For a simple illustrative example of how considering 
a deformed theory can nicely clarify aspects of the undeformed one, the reader is referred to section
1.2 of \cite{chapABA} in this collection.}

\section{Integrability and orbifolds of \texorpdfstring{$\Ncal=4$}{N=4} SYM}

Besides adding marginal operators, another way of obtaining CFT's with less supersymmetry from 
$\Ncal=4$ SYM is by orbifolding \cite{Kachru:1998ys,Lawrence:1998ja}. On the gauge theory side, this
involves picking a discrete subgroup $\Gamma$ of the $R$-symmetry group and performing the following
projection on the fields (here for $\Gamma=\Zset_M$):
\be
\phi\ra \omega^{s_\phi}\gamma \phi\gamma^{-1}\, ,\quad \text{where} \;\;
\gamma=\mathrm{diag}(1,\omega,\omega^2,\ldots, \omega^{M-1})
\;\;,\;\;\omega=e^{\frac{2\pi i}{M}}\;.
\ee
The integer $s_\phi$ is related to the $\SU(4)_R$ charge of the field $\phi$. 
The resulting theories have a quiver structure: Starting with an $\Urm(MN)$ theory, one obtains
a product gauge group $\Urm(N)_1\times\cdots
\times\Urm(N)_M$ with matter fields in bifundamental representations. The amount of supersymmetry preserved
can be $\Ncal=2$,$1$ or $0$, depending on the subgroup of $\SU(4)_R$ on which $\Gamma$ acts: $\SU(2)$, $\SU(3)$ or the whole $\SU(4)_R$
respectively. For instance, a choice of $s_\phi$ resulting in an $\Ncal=2$ theory is
$(s_X,s_Y,s_Z)=(1,-1,0)$. 

One can easily keep track of gauge invariant operators by writing them in terms of the unorbifolded fields
but with suitable phases inserted in the trace:
\be
\Tr(\gamma_m XYXZ\cdots)\,,\;\text{where}\;\; \gamma_m=\mathrm{diag}(1,\omega^m,\ldots,\omega^{(M-1)m})\,,\;m=1,\ldots,M-1\;.
\ee
Operators for different choices of $m$ do not mix with each other and correspond to different \emph{twisted 
sectors} on the string side ($m=0$ being the untwisted sector). It is easy to see that the parent and orbifolded theory 
will only differ by additional phases in the Bethe equations, as well as a 
modification of the cyclicity condition. 
The one-loop Bethe equations in various $\SU(2)$ sectors were considered in \cite{Ideguchi:2004wm}, while their
structure for the full scalar sector was derived in \cite{Beisert:2005he}, who also argued that the 
higher-loop $\Ncal=4$ SYM equations can easily be adapted to the orbifold case.\footnote{These 
authors also exhibited the Bethe equations for a \emph{combination} of orbifolding and twisting.} 
For the (X,Y) $m$--twisted $\SU(2)$ sector, the one-loop equations take the form
\be \label{OrbifoldBetheAnsatz}
e^{-\frac{4\pi i m}{M}}\left(\frac{u_k+\frac i 2}{u_k-\frac i 2}\right)^J=\prod_{j\neq k}^K\frac{u_k-u_j+i}{u_k-u_j-i}\;,
\qquad
\prod_{k=1}^K\left(\frac{u_k+\frac i2}{u_k-\frac i2}\right)=e^{\frac{2\pi i m}{M}}\;.
\ee
Note the strong similarity to the Bethe ansatz (\ref{LSBetheansatz}) for the $\beta$-deformation. 
The Bethe ansatz for more general (e.g. non-abelian) orbifolds was presented in \cite{Solovyov:2007pw}.

On the string side, one considers an $\AdS_5\times\Srm^5/\Zset_M$ background\footnote{Integrability
for $\AdS_5\times\Srm^5/\Zset_p\times \Zset_q$ orbifolds has been considered in \cite{Sadri:2005gi}.}, 
constructed via the following identifications (here for an $\Ncal=2$ orbifold):
\be
(X,Y,Z)\sim (e^{\frac{2\pi i}{M}}X,e^{-\frac{2\pi i}{M}}Y,Z) \;.
\ee
An analysis of two-spin semiclassical strings on this and more general backgrounds was 
performed in \cite{Ideguchi:2004wm} and their energies were successfully compared 
to the corresponding solutions of the orbifolded Bethe ansatz above.

An advantage of the orbifold theory compared to the parent one is that a \emph{single} giant magnon is a 
physical state. This was used in \cite{Astolfi:2007uz} to settle an issue of gauge non-invariance (dependence 
of the magnon energy on the light-cone gauge fixing parameter, once finite-size effects are
considered) which had previously arisen in the $\AdS_5\times\Srm^5$ case \cite{Arutyunov:2006gs}.
It was later argued that single magnons in $\Ncal=4$ SYM can always be thought of as living on the 
orbifolded theory \cite{Ramadanovic:2008qd}. Recently, TBA equations and wrapping effects (up to next-to-leading
order) were considered for a particular orbifold theory in \cite{Arutyunov:2010gu}.

Another interesting application of orbifold theories is that, having a new parameter $M$, one can consider 
novel scaling limits. One such limit produces the ``winding state'' \cite{Bertolini:2002nr}, where one starts with a string 
winding around an $\Srm^3/\Zset_M$ in an $\Ncal=2$ orbifold and takes $M\rightarrow\infty$ while also
taking $J$ large, keeping $M^2/J$ finite. In \cite{Astolfi:2008yw}, finite-size corrections to this state, 
as well as to orbifolded circular strings, were calculated up to order $1/J^2$ and shown to match with Bethe ansatz results.  
In a related $M\rightarrow\infty$, BMN-type limit \cite{Mukhi:2002ck}, the first finite-size corrections 
to two-impurity operators in the $\Ncal=2$ theory were computed in \cite{Astolfi:2006is}, both directly using the
dilatation operator (to two loops) and using the higher-loop version of the twisted Bethe ansatz \ref{OrbifoldBetheAnsatz}.  
They were shown to agree with each other and, given the appropriate choice of dressing factor, with the dual pp-wave string result, 
calculated using DLCQ methods (see \cite{deRisi:2004bc} for related earlier work). 

 Starting from the $\Ncal=2$ $\Urm(N)\times\Urm(N)$ quiver theory, one can 
move away from the orbifold point by varying the two gauge couplings independently, while preserving superconformal
invariance.
In \cite{Gadde:2010zi} this was shown to break integrability, but in the extremal case where one of the two couplings 
vanishes (and we obtain an $\SU(N)$ gauge theory with $N_f=2N$ flavors) it appeared that integrability might be recovered.
This result, if confirmed, would provide a first example of an integrable theory in the Veneziano limit 
($N,N_f\ra\infty$ with $N/N_f$ constant) instead of the usual 't Hooft limit. Recently,
\cite{Gadde:2010ku} considered magnon propagation on such interpolating non-integrable chains.

On the amplitude side, it is known that orbifold theories are planar 
equivalent to the parent theory to all orders in perturbation theory \cite{Bershadsky:1998cb}. Thus the BDS 
iterative conjecture is expected to immediately transfer to the orbifold theories.

\subsection{Other backgrounds}

 Apart from the orbifold theories discussed above, there exist several AdS/CFT setups with 
reduced supersymmetry in the literature, and one can ask whether integrability
appears in those cases as well. Perhaps the best-known example of this kind \cite{Klebanov:1998hh}
is constructed by taking the near horizon limit of a stack of D-branes situated at the tip
of the \emph{conifold}, a noncompact 6-dimensional Calabi--Yau manifold which can be written
as a cone over the 5-dimensional \emph{Sasaki--Einstein} manifold known as $T^{1,1}$.
The near horizon geometry of this system is $\AdS_5\times T^{1,1}$ and corresponds to an 
$\Ncal=1$ superconformal $\Urm(N)\times\Urm(N)$ gauge theory with bifundamentals, which
is an infrared limit of a $\Zset_2$ orbifold theory of the type discussed above. 

There has been intense activity in constructing semiclassical string solutions on $T^{1,1}$,
as well as generalisations known as $T^{p,q},Y^{p,q}$ and $L^{p,q,r}$
\cite{Schvellinger:2003vz,Kim:2003vn,Wang:2005baa,Benvenuti:2005cz,Benvenuti:2008bd,Giataganas:2009dr,
Dimov:2009ut,Giataganas:2010mj}. 
However, these conformal fixed points only appear at strong coupling, and thus do not correspond
to perturbatively finite field theories. It is therefore far from obvious that one should expect 
to find integrability. Indeed, no Lax pair construction is known for these backgrounds. 
Furthermore, as observed in \cite{Dimov:2009ut} for $T^{1,1}$ and its $\beta$--deformed analogue,
the dispersion relation for magnons and spiky strings is transcendental, in stark contrast to the 
$\AdS_5\times \Srm^5$ case. This is a clear indication that integrability, if it appears at all,  
would have to do so in a much more complicated way than in $\Ncal=4$ SYM. 
On the other hand, it was shown in \cite{Benvenuti:2007qt} that, for the cases
mentioned above, the bosonic sector in the near-flat-space limit \cite{Maldacena:2006rv} 
is the same as for $\Srm^5$. Thus the full sigma models do at least possess an integrable subsector.

\section{Open spin chain boundary conditions}

 One can also investigate integrability in a less supersymmetric setting by considering
systems involving spin chains with \emph{open} boundary conditions. This clearly signals the presence 
of open strings, and therefore D-branes, on the dual string side. After reviewing some universal 
aspects of open spin chains, we will proceed to discuss several different situations where they 
make an appearance in the AdS/CFT context. 

As  reviewed in \cite{chapABA} in this collection, in the algebraic approach to integrability for
closed spin chains one begins by considering the RTT relations for the monodromy matrix, defined 
in terms of an $R$-matrix satisfying the Yang--Baxter equation:
\be
R_{12}(u,v)R_{13}(u,w)R_{23}(v,w)=R_{23}(v,w)R_{13}(u,w)R_{12}(u,v)\;.
\ee
For open chains, these equations still hold, but have to be supplemented (at each boundary)
with the \emph{reflection}, or \emph{boundary Yang--Baxter} equation 
\cite{Sklyanin:1988yz}:
\be
R_{12}(u,v)K_1(u)R_{21}(v,-u)K_2(v)=K_2(v)R_{12}(u,-v)K_1(u)R_{21}(-u,-v)\;.
\ee
Here the $K_{1,2}(u)$ are known as the boundary reflection matrices. See e.g. \cite{Arnaudon:2004sd}
for a discussion of various boundary conditions, and the corresponding reflection matrices,
for $\mathfrak{sl}(n)$ and $\mathfrak{sl}(m|n)$ spin chains, as well as further references to the open-chain
literature. In the special case where the boundary 
conditions preserve the same $\mathfrak{gl}(n)$ symmetry as the bulk chain (which is often not the case 
in the setups to be considered below), the general form of a perturbatively long-range
integrable $\mathfrak{gl}(n)$ spin chain with open boundary conditions was given in \cite{Beisert:2008cf}.


The generic structure of  any putative open string Bethe ansatz is
\be \label{OpenBA}
e^{2ip_kL}=B_1(p_k)B_2(p_k)\prod_{j=1,j\neq k}^{M} S_{jk}(p_j,p_k)S_{kj}(-p_j,p_k)
\ee
where the $S_{jk}$ are the bulk S-matrices, and $B_{1,2}$ are the boundary reflection matrices.
To understand the above structure (see also \cite{Beisert:2008cf} for a nice exposition), 
note that a given excitation moving with momentum $p_i$ 
will scatter with a number of other excitations, reflect
from the boundary, scatter with the other excitations again (but with opposite momentum) and 
reflect from the second boundary before finally returning to its original position.
Assuming that the bulk theory is integrable, the question of integrability 
hinges on the precise form of the boundary matrices $B_{1,2}$.

In the closed-chain case the Bethe ansatz is normally accompanied by a cyclicity condition (which 
on the string side arises from the closed-string level-matching condition). However, this is absent 
for the open-chain case. An immediate consequence of this is that single-impurity states are physical,
even for non-zero momentum. 

As in the closed spin-chain case, new effects arise when considering long-range \emph{short} open spin 
chains, in particular \emph{spanning} terms, which are the
analogues of the closed-chain wrapping interactions for finite-length open spin chains. Little is known
at present about their structure from the field theory side, though a study of such terms in
\cite{Beisert:2008cf} suggests that they are not strongly
constrained by integrability, which would therefore appear to lose some of its predictive power for short chains.

\subsection{Open spin chains within \texorpdfstring{$\Ncal=4$}{N=4} SYM}

 Although  this review is mainly concerned with integrable theories beyond $\Ncal=4$ SYM, 
there exist several interesting cases where integrable open spin chains arise \emph{within} the $\Ncal=4$ theory
itself. We will thus first discuss this class of theories, which arise through the consideration 
of non-trivial backgrounds within $\Ncal=4$ SYM.

\subsubsection{Open strings on giant gravitons}

The first case of this type is that of open strings ending on maximal giant gravitons \cite{McGreevy:2000cw}
in $\AdS_5\times\Srm^5$. These are $D3$-branes wrapping 3-cycles inside the
5-sphere. The gauge theory picture is that 
of an open-spin chain word attached to a baryon-like (determinant) operator in $\Ncal=4$ SYM,
formed out of one of the scalars in the theory, here denoted $\Phi_B$: 
\be
\epsilon_{i_1\cdots i_N}\epsilon^{j_1\cdots j_N}(\Phi_B)^{i_1}_{j_1}\cdots (\Phi_B)^{i_{N-1}}_{j_{N-1}} 
(\Phi_{k_1}\Phi_{k_2}\cdots\Phi_{k_L})^{i_N}_{\;j_N}\;.
\ee
In the large-$N$ limit the determinant part becomes very heavy and has no dynamics of its own,
so this system behaves as a spin chain with open boundary conditions. 

The one-loop hamiltonian for this type of chain was considered in
\cite{Berenstein:2005vf} and shown to be integrable.  It was further investigated  
at two-loops in \cite{Agarwal:2006gc}, with the final two-loop result, in the $\SU(2)$ sector, 
given in \cite{Hofman:2007xp}:
\be \label{MGGhamiltonian}
H=\!(2g^2-8g^4)\sum_{i=1}^\infty\left(I\!-P_{i,i+1}\right)
+2g^4\sum_{i=1}^\infty \left(I\!-P_{i,i+2}\right)
+(2g^2-4g^4)q_1^{\Phi_B}+2g^4 q_2^{\Phi_B}
\ee
with $q_i^{\Phi_B}=1$ if $\Phi_{i}=\Phi_B$ and 0 otherwise. 
The first two terms are the same as the bulk hamiltonian,
the third is the naive boundary contribution coming from all
the derivatives in the dilatation operator acting outside the
determinant, while the last term comes from one of the derivatives
acting \emph{on} the determinant. This term is naively $1/N$ suppressed, but survives in the 
planar limit, the suppression being compensated by the fact that it can act
on any of the $N-1$ terms in the determinant. As shown in \cite{Hofman:2006xt}, the hamiltonian 
(\ref{MGGhamiltonian}) is consistent with integrability.
On the string side, \cite{Mann:2006rh} constructed non-local conserved sigma-model charges for 
classical open strings ending on maximal giant gravitons in the full bosonic sector, thus
providing strong supporting evidence for all-loop integrability of the maximal graviton system.

For non-maximal giant gravitons (which correspond to sub-determinant-type operators
in the gauge theory) the open spin chain becomes dynamical, in the sense that 
the number of sites can vary, even at one-loop level.
This case was investigated in \cite{Berenstein:2005fa,Berenstein:2006qk}, where it was argued that
the formalism of \emph{Cuntz chains} provides a better description than the standard spin chain, and
some (numerical) evidence for integrability was provided. However, on the string side, the
appearance of extra conditions  hinders the construction of non-local 
sigma model charges \cite{Mann:2006rh}. Thus the prospects for integrability in this case do not look 
particularly good.\footnote{Nevertheless, integrability was recently demonstrated for giant magnons scattering 
off $Y=0$ non-maximal gravitons \cite{Ciavarella:2010tp}, indicating that integrable subcases do exist.}

\mparagraph{Reflecting magnons} 

Giant magnons ending on maximal giant gravitons were considered in \cite{Hofman:2007xp}. 
One can, without loss of generality, choose to consider open chains made up of a large
number of $Z$ fields, which, on the string side, correspond to semiclassical strings with
a large angular momentum along the $5-6$ plane within $\Srm^5$. One can then consider
two different orientations of the giant magnon relative to this plane.

\vspace{.1cm}
{\flushleft \it The $Y=0$ magnon}:
In this case we choose the $D3$-brane to wrap the 3-sphere defined by $Y=0$, which 
corresponds to the operator $\det Y$ in the gauge theory. Attaching an open spin chain to this
determinant, we are led to an operator of the form:
\be
\epsilon_{i_1\cdots i_N}\epsilon^{j_1\cdots j_N}Y^{i_1}_{j_1}\cdots Y^{i_{N-1}}_{j_{N-1}}
\left(Z\cdots Z\chi Z \cdots Z\right)^{i_N}_{j_N} \;.
\ee
Here $\chi$ stands for any impurity, though it will need to be a $Y$ field if we wish to stay within
the $\SU(2)$ sector. As explained in \cite{Hofman:2007xp}, this configuration has no boundary 
degrees of freedom, and there is a unique vacuum state. 
The boundary preserves an $\SU(1|2)^2$ out of the bulk $\SU(2|2)$ symmetry. 
The boundary scattering phase was found in \cite{Chen:2007ec}, while commuting open-chain transfer matrices, 
necessary for the construction of the Bethe ansatz,
were derived in \cite{Murgan:2008fs}.\footnote{The works \cite{Murgan:2008zu,Murgan:2009bx} 
generalised the $q$-deformed S-matrix of \cite{Beisert:2008tw} to the $Y=0$ and $Z=0$ magnon context, 
and studied open--chain transfer matrices for these cases.} In \cite{Ahn:2010xa} it was shown that part
of the bulk Yangian symmetry persists for boundary scattering and can be used to determine the bound-state 
reflection matrices. This boundary Yangian was further discussed in \cite{MacKay:2010ey}. 
The higher--loop Bethe ansatz for this class of operators was proposed in \cite{Galleas:2009ye},
see also \cite{Okamura:2006zr} for an earlier discussion. A different derivation, which agrees
with the one above, is in \cite{Nepomechie:2009zi}. 

\vspace{.1cm}
{\flushleft \it The $Z=0$ magnon}:
Here we consider a $D3$-brane wrapping the 3-sphere defined by $Z=0$, which is dual to
the gauge theory operator $\det Z$. The open chain is still made up mainly of $Z$'s, but
it is easy to see that they cannot be attached directly to the determinant: Such a configuration
would factorise into a determinant plus a trace. To obtain a nontrivial open spin chain, there need to 
be impurities (fields other than $Z$) stuck to the boundary:\footnote{In the $\SU(2)$ sector,
all the $\chi$'s will have to be of the same type, e.g. $Y$ fields.} 
\be
\epsilon_{i_1\cdots i_N}\epsilon^{j_1\cdots j_N}Z^{i_1}_{j_1}\cdots Z^{i_{N-1}}_{j_{N-1}}
\left(\chi Z\cdots Z\chi'' Z \cdots Z\chi'\right)^{i_N}_{j_N} \;.
\ee
In this case there \emph{are} boundary degrees of freedom, which (like the bulk magnon) fall into
representations of $\SU(2|2)^2$ \cite{Hofman:2007xp}. There are thus 16 states living on each boundary, 
which were identified on the string side in \cite{Bak:2008xq}, by considering fermionic zero modes around the
finite-size string solution for an open string ending on the $Z=0$ graviton.\footnote{The string solution
itself was previously found in unpublished work by C.~Ahn, D.~Bak and S.J.~Rey.} 
The boundary scattering phase was found in \cite{Ahn:2007bq}.
One notable feature of the $Z=0$ case is the presence of poles in the reflection amplitude not
corresponding to bound states, whose origin was explained in \cite{Palla:2008zc}.
As for $Y=0$, a boundary $R$-matrix was proposed in \cite{Hofman:2007xp}, however it
did not directly satisfy the BYBE. This was reconsidered in \cite{Ahn:2008df}, who found a suitable 
basis where the boundary $R$ matrix does satisfy the BYBE. 
The higher-loop nested Bethe ansatz in this case was constructed in \cite{Correa:2009dm}.

\mparagraph{Finite-size effects}

Considerable recent activity in the $\Ncal=4$ SYM context has centered around understanding
finite-size effects, or wrapping interactions on the gauge theory side (see the 
reviews \cite{chapLuescher,chapHigher,chapTrans,chapTBA} in this collection). There is an analogous formalism
for the open-chain case, which was used in \cite{Correa:2009mz} to compute
L\"uscher-type corrections to open strings ending on giant gravitons (for vacuum states) and compare with explicit 
gauge theory results. The anomalous dimension of the $Y=0$ vacuum chain was shown to vanish (a result expected
by supersymmetry) while in the $Z=0$ case it was non-trivial. The L\"uscher formulae of \cite{Correa:2009mz} were
extended to the multiparticle case in \cite{Bajnok:2010ui}, allowing the computation of finite-size corrections
to one-excitation states in the $Y=0$ case and leading to an explicit prediction to be checked by future gauge theory 
perturbative calculations. The analogous computation for the (more challenging) $Z=0$ brane has not yet been performed. 
Furthermore, no TBA or Y-system equations are available at present for the boundary case. 

Classical solutions for finite-size magnons on $Z=0$ gravitons (generalising those in \cite{Bak:2008xq}) 
can be found in \cite{Ciavarella:2010je}.

\mparagraph{Other graviton-magnon combinations}\mbox{}

The work \cite{Correa:2006yu} studied open strings ending on giant gravitons in the AdS part of 
the geometry and, on the gauge side, identified the planar dilatation operator as the hamiltonian 
of an open $sl(2)$ spin chain. However, novel features such as a variable occupation number and 
continuous bands in the spectrum prevented a clear understanding of integrability in this case.
Other configurations of strings on giant gravitons have been considered in \cite{Sadri:2003mx} (in 
the BMN limit), as well as in \cite{Hirano:2006ti}, where gauge theory operators 
dual to a giant graviton/magnon bound state are proposed.

\subsubsection{Operators with very large R-charge}

Giant gravitons are dual to baryonic operators in $\Ncal=4$ SYM whose
dimensions grow linearly with $N$. One can consider other types of operators
whose dimension grows as $N^2$, which in the simplest case
are of the form $(\det Z)^M$ (with $M\sim N$) but more generally are described by Schur polynomial 
operators related to the Young diagram encoding their symmetries. On the dual gravity side the
number of $D3$-branes is so large that it is no longer possible to ignore backreaction, 
and this modifies the $\AdS$ geometry into an LLM-type background.   
Strings ``attached'' to the above operators\footnote{Note that these are
actually closed strings, since after the $D3$-branes
have backreacted there are no explicit open strings on the background.}
 have recently been considered from the gauge theory side in \cite{deMelloKoch:2009zm}.
It is possible to integrate out the effect of the background and construct an
effective dilatation operator, which is integrable in a certain limit. Interestingly,
this limit includes non-planar diagrams between the trace operator and the background.
Although, as reviewed in \cite{chapObserv},  truly non-planar contributions (acting on the trace operator 
by splitting and joining) are still expected to spoil integrability, this novel integrable 
limit of $\Ncal=4$ SYM is still interesting and deserves further exploration.

\subsubsection{Open string insertions on Wilson loops}

In the absence of nontrivial background operators for the spin chain to end on, open 
string boundary conditions would not be gauge invariant. A way to avoid this problem
is to consider open chain insertions on Wilson loops \cite{Drukker:2006xg}. As shown
in that work, which considered such operators in the $\SU(2)$ sector at one loop,
the boundary conditions turn out to be purely reflective (Neumann). Thus the Bethe
ansatz can be related to a closed-chain one by the method of images. The dual description
of the Wilson loop (which has  angular momenta on $\Srm^5$ to account for the scalar 
insertions) was shown to reduce to ``half'' the standard closed folded string solution, whose
energy precisely matches the Bethe ansatz computation. This setup is thus at least one-loop 
integrable (no higher-loop checks have been performed at present).

\subsection{Theories with fundamental flavor}

One can also obtain open spin chains by extending the field content of $\Ncal=4$ SYM by adding
flavors, i.e. fields in the fundamental representation of the gauge group. Introducing such fields in
the spectrum means that, apart from trace operators, one can construct gauge--invariant operators 
of the generic form:
\be
\bar{Q}\Phi_{i_1}\Phi_{i_2}\cdots \Phi_{i_L} Q
\ee
where $Q$ is one of the fundamental fields. This operator, having no cyclicity properties,
will behave as an open chain. We will now review three distinct settings where these types of operators have 
been studied in an integrability context.

\subsubsection{The orientifold theory}

In this setup, one considers a D3--O7--D7 system, where one first performs an orientifold projection
and then adds the required number of D7 branes (four, plus their mirrors) to cancel the orientifold charge.
The result is $\Ncal=2$ SYM with gauge group $\mathrm{Sp}(N)$, one hypermultiplet in the antisymmetric representation
and four in the fundamental,
which is known to be a finite theory.\footnote{A different type of orientifold
which preserves $\Ncal=4$ SYM but leads to gauge group $\SO(N)$ or $\mathrm{Sp}(N)$ was recently considered
in \cite{Caputa:2010ep}, though in that case the focus was on non-planar corrections, the differences 
to $\SU(N)$  being relatively minor at planar level.}  The $\Ncal=2$ vector multiplet contains an adjoint chiral multiplet
$W$, while the antisymmetric hypermultiplet two chiral multiplets $Z,Z'$. 
The near-horizon geometry is that of
an $\AdS_5\times \Srm^5/\Zset_2$ orientifold. Here the $\Zset_2$ acts as $Z\rightarrow -Z$ (or
$\varphi_3\rightarrow \varphi_3+\pi$),
leaving a fixed plane at $Z=0$. The worldsheet coordinate is also identified  
as $\sigma\rightarrow \pi-\sigma$. 

Relatively few studies of integrability have been undertaken for this theory. The pp-wave
spectrum was discussed in \cite{Berenstein:2002zw}. 
Several open spinning string solutions on the dual orientifold were considered in \cite{Stefanski:2003qr}.  
In \cite{Chen:2004mu}, the one-loop hamiltonian for the $\SU(3)$ sector comprised of $W,Z,Z'$ was shown to be integrable 
and the corresponding one-loop Bethe ansatz constructed. In the  $(Z,Z')$ $\SU(2)$ sector, it is:
\be \label{OrientBA}
\left(\frac{u_k+\frac i2}{u_k-\frac i2}\right)^{2L}=\prod_{j\neq k}^{K}
\frac{u_k-u_j+i}{u_k-u_j-i}\frac{u_k+u_j+i}{u_k+u_j-i}
\ee
Notice that it is of the form (\ref{OpenBA}). Applying the doubling trick, by means of which this 
Bethe ansatz can be related to a closed string one with the extra condition
that the set of roots is symmetric under $u_j\ra -u_j$, energies of two-spin open strings were successfully
compared to gauge theory in \cite{Chen:2004yf}. At the time of writing three-spin strings have not been compared, 
while the question of higher-loop integrability is still open.

\subsubsection{The D3--D7--brane system}

Here one considers $\AdS_5\times \Srm^5$, with a D7-brane filling $\AdS_5$ and wrapping 
an $\Srm^3$ in $\Srm^5$. Unlike the case above, this theory is conformal only in the strict large-$N$
limit, where the backreaction of the D7 brane can be ignored. On the gauge theory side, this corresponds
to ignoring $1/N$-suppressed processes with virtual fundamental flavors 
between bulk states (which would provide a non-zero contribution to the $\beta$-function).  

The bulk hamiltonian is the same as for $\Ncal=4$ SYM, so closed spin chains in this setup are automatically
integrable. The one-loop open-chain hamiltonian is integrable as well, with trivial 
boundary terms \cite{Erler:2005nr}. The one-loop, $\SU(2)$-sector Bethe ansatz is precisely the same as (\ref{OrientBA}).
The higher-loop reflection matrices for this case were studied in \cite{Correa:2008av}, where it
was shown that integrability survives, largely thanks to the fact that the boundary respects the  
$\mathfrak{psl}(2|2)\times \overline{\mathfrak{psl}(2|2)}$ factorisation of the bulk theory. More recently,
the work \cite{MacKay:2010zb} extended these results by constructing the reflection matrices for boundary 
scattering of \emph{bound states}. 

 On the gravity side, \cite{Mann:2006rh} showed integrability for the full bosonic sector by
observing that the equations governing open string motion are practically the same as in 
the maximal giant graviton case discussed above. It is thus expected that this system exhibits 
higher-loop integrability.

\subsubsection{Defect theories}

A different setup with fundamentals can be obtained by considering a D3--D5 system, with
a single D5 sharing only three directions (say $x^0,x^1$ and $x^2$) with the stack 
of $N$ D3 branes. The configuration thus has four Neumann--Dirichlet directions and
preserves supersymmetry. Taking the D3-brane near-horizon limit, we obtain the usual
$\AdS_5\times \Srm^5$ geometry, but now the D5 brane wraps an $\AdS_4\times \Srm^2$ in
$\AdS_5\times\Srm^5$. On the gauge theory side, we obtain $\Ncal=4$ SYM coupled to a 
\emph{defect} located at $x^3=0$. The matter content on the defect is a 3d $\SU(N)$ vector 
multiplet plus a 3d fundamental hypermultiplet (containing two chiral multiplets $q_{1,2}$).

As shown in \cite{DeWolfe:2004zt}, starting from a ground state of the form $\bar{q}_1 Z\cdots Z q_1$
there are two types of excitations one can consider: If the excitations are along the D5 brane, 
the boundary conditions are Dirichlet, which on the gauge theory translates to the boundary
term being fixed. Otherwise, the string satisfies Neumann 
boundary conditions, which for the spin chain means that the boundary excitations are dynamical:
The boundary state can flip from $q_1$ to $q_2$, which effectively increases the length of the chain by 1. 
In both cases the boundary matrix is trivial and the full bosonic sector is integrable at one loop. 
As before, there is no boundary phase in the $\SU(2)$ sector, though it does make an appearance in the $\SL(2)$ sector
\cite{McLoughlin:2005gj}. Spinning string solutions in this setup were considered 
in \cite{Susaki:2004tg,Susaki:2005qn,Okamura:2005cj}.

However, it was eventually understood that this one-loop integrability is an accident. The 
first indication came from the gravity side, when
\cite{Mann:2006rh} showed that nonlocal charges could only be constructed in the $\SU(2)$ sector.
Finally, by careful analysis of the symmetries, \cite{Correa:2008av} constructed the all-loop
reflection matrices (aspects of which were previously considered in \cite{Okamura:2006zr})
with the result that they do \emph{not} satisfy the BYBE. 

\section{Outlook}

In this short review we gave an overview of several different known ways of pushing integrability
beyond the highly symmetric case of $\Ncal=4$ SYM. As we have seen, it is relatively easy to maintain
integrability at the one-loop level in less supersymmetric (but still superconformal) 
situations, but all-loop integrability is a much more stringent requirement. 
Indeed, it appears that all non-$\Ncal=4$ SYM models where higher-loop integrability persists are really just 
$\Ncal=4$ SYM in disguise, in the sense that the bulk spin chain is undeformed, with 
differences arising only in the boundary conditions: Twisted ones for the real--$\beta$ deformations, 
orbifold ones for the quiver theories, and open ones for giant gravitons  
and theories with fundamentals.

This observation seems to reaffirm how special the  $\Ncal=4$ SYM theory is, even within the already 
very restricted class of superconformal quantum field theories. On the other hand, 
the rich pattern of integrability breaking in the theories discussed above should help us
better appreciate the implications (and limitations) of integrability for more realistic
theories, in a more controllable setting than that of QCD. Even in those cases which \emph{are} believed 
to be higher-loop integrable, there remain numerous open questions whose resolution can be expected to
contribute to a deeper understanding of AdS/CFT integrability, and ultimately of the AdS/CFT
correspondence itself.

\paragraph{Acknowledgments} I am grateful to N.\ Beisert, R.\ Janik, C.\ Kristjansen, T.\ M\aa nsson, C.\ Sieg 
and A.\ Torrielli for useful discussions, comments and clarifications during the writing of 
this review. This work was supported by FNU through grant number 272-08-0329.

\phantomsection
\addcontentsline{toc}{section}{\refname}
\bibliography{chapters,ChapIV2ref}

\begin{thebibliography}{100}
\ifx\href\asklfhas\newcommand{\href}[2]{#2}\fi
\ifx\arxivref\asklfhas\newcommand{\arxivref}[2]{\href{http://arxiv.org/abs/#1}%
{#2}}\fi
\ifx\doiref\asklfhas\newcommand{\doiref}[2]{\href{http://dx.doi.org/#1}{#2}}\fi
\raggedright
\small
\parskip 0pt

\bibitem{chapQCD}
G.~Korchemsky,
\textit{``Review of AdS/CFT Integrability, Chapter IV.4: Integrability in QCD
  and $\mathcal{N}$ $<$ 4 SYM''},
\texttt{\arxivref{1012.4000}{arxiv:1012.4000}}.

\bibitem{chapN6}
T.~Klose,
\textit{``Review of AdS/CFT Integrability, Chapter IV.3: $\mathcal{N}$ = 6
  Chern-Simons and Strings on $AdS_4 \times CP^3$''},
\texttt{\arxivref{1012.3999}{arxiv:1012.3999}}.

\bibitem{Leigh:1995ep}
R.~G.~Leigh and M.~J.~Strassler,
\textit{``{Exactly marginal operators and duality in four-dimensional N=1
  supersymmetric gauge theory}''},
\textsf{\doiref{10.1016/0550-3213(95)00261-P}{Nucl.~Phys.~B447,~95~(1995)}},
\texttt{\arxivref{hep-th/9503121}{hep-th/9503121}}.

\bibitem{Aharony:2002tp}
O.~Aharony and S.~S.~Razamat,
\textit{``{Exactly marginal deformations of N = 4 SYM and of its supersymmetric
  orbifold descendants}''},
\textsf{\doiref{10.1088/1126-6708/2002/05/029}{JHEP~0205,~029~(2002)}},
\texttt{\arxivref{hep-th/0204045}{hep-th/0204045}}.

\bibitem{Mauri:2005pa}
A.~Mauri, S.~Penati, A.~Santambrogio and D.~Zanon,
\textit{``{Exact results in planar N = 1 superconformal Yang-Mills theory}''},
\textsf{\doiref{10.1088/1126-6708/2005/11/024}{JHEP~0511,~024~(2005)}},
\texttt{\arxivref{hep-th/0507282}{hep-th/0507282}}.

\bibitem{Lunin:2005jy}
O.~Lunin and J.~M.~Maldacena,
\textit{``{Deforming field theories with U(1) x U(1) global symmetry and their
  gravity duals}''},
\textsf{\doiref{10.1088/1126-6708/2005/05/033}{JHEP~0505,~033~(2005)}},
\texttt{\arxivref{hep-th/0502086}{hep-th/0502086}}.

\bibitem{Frolov:2005dj}
S.~Frolov,
\textit{``{Lax pair for strings in Lunin-Maldacena background}''},
\textsf{\doiref{10.1088/1126-6708/2005/05/069}{JHEP~0505,~069~(2005)}},
\texttt{\arxivref{hep-th/0503201}{hep-th/0503201}}.

\bibitem{Berenstein:2000ux}
D.~Berenstein, V.~Jejjala and R.~G.~Leigh,
\textit{``{Marginal and relevant deformations of N = 4 field theories and
  non-commutative moduli spaces of vacua}''},
\textsf{\doiref{10.1016/S0550-3213(00)00394-1}{Nucl.~Phys.~B589,~196~(2000)}},
\texttt{\arxivref{hep-th/0005087}{hep-th/0005087}}.

\bibitem{Roiban:2003dw}
R.~Roiban,
\textit{``{On spin chains and field theories}''},
\textsf{\doiref{10.1088/1126-6708/2004/09/023}{JHEP~0409,~023~(2004)}},
\texttt{\arxivref{hep-th/0312218}{hep-th/0312218}}.

\bibitem{Berenstein:2004ys}
D.~Berenstein and S.~A.~Cherkis,
\textit{``{Deformations of N = 4 SYM and integrable spin chain models}''},
\textsf{\doiref{10.1016/j.nuclphysb.2004.09.005}{Nucl.~Phys.~B702,~49~(2004)}},
\texttt{\arxivref{hep-th/0405215}{hep-th/0405215}}.

\bibitem{Beisert:2005if}
N.~Beisert and R.~Roiban,
\textit{``{Beauty and the twist: The Bethe ansatz for twisted N = 4 SYM}''},
\textsf{\doiref{10.1088/1126-6708/2005/08/039}{JHEP~0508,~039~(2005)}},
\texttt{\arxivref{hep-th/0505187}{hep-th/0505187}}.

\bibitem{chapYang}
A.~Torrielli,
\textit{``Review of AdS/CFT Integrability, Chapter VI.2: Yangian Algebra''},
\texttt{\arxivref{1012.4005}{arxiv:1012.4005}}.

\bibitem{Ihry:2008gm}
J.~N.~Ihry,
\textit{``{Yangians in Deformed Super Yang-Mills Theories}''},
\textsf{\doiref{10.1088/1126-6708/2008/04/051}{JHEP~0804,~051~(2008)}},
\texttt{\arxivref{0802.3644}{arxiv:0802.3644}}.

\bibitem{Frolov:2005ty}
S.~A.~Frolov, R.~Roiban and A.~A.~Tseytlin,
\textit{``{Gauge - string duality for superconformal deformations of N = 4
  super Yang-Mills theory}''},
\textsf{\doiref{10.1088/1126-6708/2005/07/045}{JHEP~0507,~045~(2005)}},
\texttt{\arxivref{hep-th/0503192}{hep-th/0503192}}.

\bibitem{Ahn:2010ws}
C.~Ahn, Z.~Bajnok, D.~Bombardelli and R.~I.~Nepomechie,
\textit{``{Twisted Bethe equations from a twisted S-matrix}''},
\textsf{\doiref{10.1007/JHEP02(2011)027}{JHEP~1102,~027~(2011)}},
\texttt{\arxivref{1010.3229}{arxiv:1010.3229}}.

\bibitem{Grassi:2006tj}
P.~A.~Grassi and J.~Kluson,
\textit{``{Pure spinor strings in TsT deformed background}''},
\textsf{\doiref{10.1088/1126-6708/2007/03/033}{JHEP~0703,~033~(2007)}},
\texttt{\arxivref{hep-th/0611151}{hep-th/0611151}}.

\bibitem{chapSpinning}
A.~Tseytlin,
\textit{``Review of AdS/CFT Integrability, Chapter II.1: Classical $AdS_5\times
  S^5$ string solutions''},
\texttt{\arxivref{1012.3986}{arxiv:1012.3986}}.

\bibitem{chapQString}
T.~McLoughlin,
\textit{``Review of AdS/CFT Integrability, Chapter II.2: Quantum Strings in
  $AdS_5\times S^5$''},
\texttt{\arxivref{1012.3987}{arxiv:1012.3987}}.

\bibitem{Chen:2005sba}
H.-Y.~Chen and S.~Prem~Kumar,
\textit{``{Precision test of AdS/CFT in Lunin-Maldacena background}''},
\textsf{\doiref{10.1088/1126-6708/2006/03/051}{JHEP~0603,~051~(2006)}},
\texttt{\arxivref{hep-th/0511164}{hep-th/0511164}}.

\bibitem{Chen:2006bh}
H.-Y.~Chen and K.~Okamura,
\textit{``{The anatomy of gauge / string duality in Lunin-Maldacena
  background}''},
\textsf{\doiref{10.1088/1126-6708/2006/02/054}{JHEP~0602,~054~(2006)}},
\texttt{\arxivref{hep-th/0601109}{hep-th/0601109}}.

\bibitem{Hofman:2006xt}
D.~M.~Hofman and J.~M.~Maldacena,
\textit{``{Giant magnons}''},
\textsf{\doiref{10.1088/0305-4470/39/41/S17}{J.~Phys.~A39,~13095~(2006)}},
\texttt{\arxivref{hep-th/0604135}{hep-th/0604135}}.

\bibitem{Chu:2006ae}
C.-S.~Chu, G.~Georgiou and V.~V.~Khoze,
\textit{``{Magnons, classical strings and beta-deformations}''},
\textsf{\doiref{10.1088/1126-6708/2006/11/093}{JHEP~0611,~093~(2006)}},
\texttt{\arxivref{hep-th/0606220}{hep-th/0606220}}.

\bibitem{Bobev:2005cz}
N.~P.~Bobev, H.~Dimov and R.~C.~Rashkov,
\textit{``{Semiclassical strings in Lunin-Maldacena background}''},
\texttt{\arxivref{hep-th/0506063}{hep-th/0506063}}.

\bibitem{Bobev:2007bm}
N.~P.~Bobev and R.~C.~Rashkov,
\textit{``{Spiky Strings, Giant Magnons and beta-deformations}''},
\textsf{\doiref{10.1103/PhysRevD.76.046008}{Phys.~Rev.~D76,~046008~(2007)}},
\texttt{\arxivref{0706.0442}{arxiv:0706.0442}}.

\bibitem{Bykov:2008bj}
D.~V.~Bykov and S.~Frolov,
\textit{``{Giant magnons in TsT-transformed $AdS_5 x S^5$}''},
\textsf{\doiref{10.1088/1126-6708/2008/07/071}{JHEP~0807,~071~(2008)}},
\texttt{\arxivref{0805.1070}{arxiv:0805.1070}}.

\bibitem{Ahn:2010da}
C.~Ahn and P.~Bozhilov,
\textit{``{Finite-Size Dyonic Giant Magnons in TsT-transformed $AdS_5\times
  S^5$}''},
\textsf{\doiref{10.1007/JHEP07(2010)048}{JHEP~1007,~048~(2010)}},
\texttt{\arxivref{1005.2508}{arxiv:1005.2508}}.

\bibitem{chapLuescher}
R.~Janik,
\textit{``Review of AdS/CFT Integrability, Chapter III.5: L\"uscher
  corrections''},
\texttt{\arxivref{1012.3994}{arxiv:1012.3994}}.

\bibitem{chapTBA}
Z.~Bajnok,
\textit{``Review of AdS/CFT Integrability, Chapter III.6: Thermodynamic Bethe
  Ansatz''},
\texttt{\arxivref{1012.3995}{arxiv:1012.3995}}.

\bibitem{chapTrans}
V.~Kazakov and N.~Gromov,
\textit{``Review of AdS/CFT Integrability, Chapter III.7: Hirota Dynamics for
  Quantum Integrability''},
\texttt{\arxivref{1012.3996}{arxiv:1012.3996}}.

\bibitem{Gromov:2010dy}
N.~Gromov and F.~Levkovich-Maslyuk,
\textit{``{Y-system and beta-deformed N=4 Super-Yang-Mills}''},
\textsf{\doiref{10.1088/1751-8113/44/1/015402}{J.~Phys.~A44,~015402~(2011)}},
\texttt{\arxivref{1006.5438}{arxiv:1006.5438}}.

\bibitem{Freedman:2005cg}
D.~Z.~Freedman and U.~Gursoy,
\textit{``{Comments on the beta-deformed N = 4 SYM theory}''},
\textsf{\doiref{10.1088/1126-6708/2005/11/042}{JHEP~0511,~042~(2005)}},
\texttt{\arxivref{hep-th/0506128}{hep-th/0506128}}.

\bibitem{chapHigher}
C.~Sieg,
\textit{``Review of AdS/CFT Integrability, Chapter I.2: The spectrum from
  perturbative gauge theory''},
\texttt{\arxivref{1012.3984}{arxiv:1012.3984}}.

\bibitem{Fiamberti:2010fw}
F.~Fiamberti, A.~Santambrogio and C.~Sieg,
\textit{``{Superspace methods for the computation of wrapping effects in the
  standard and beta-deformed N=4 SYM}''},
\texttt{\arxivref{1006.3475}{arxiv:1006.3475}}.

\bibitem{Fiamberti:2008sm}
F.~Fiamberti, A.~Santambrogio, C.~Sieg and D.~Zanon,
\textit{``{Finite-size effects in the superconformal beta-deformed N=4 SYM}''},
\textsf{\doiref{10.1088/1126-6708/2008/08/057}{JHEP~0808,~057~(2008)}},
\texttt{\arxivref{0806.2103}{arxiv:0806.2103}}.

\bibitem{Fiamberti:2008sn}
F.~Fiamberti, A.~Santambrogio, C.~Sieg and D.~Zanon,
\textit{``{Single impurity operators at critical wrapping order in the
  beta-deformed N=4 SYM}''},
\textsf{\doiref{10.1088/1126-6708/2009/08/034}{JHEP~0908,~034~(2009)}},
\texttt{\arxivref{0811.4594}{arxiv:0811.4594}}.

\bibitem{Arutyunov:2010gu}
G.~Arutyunov, M.~de~Leeuw and S.~J.~van~Tongeren,
\textit{``{Twisting the Mirror TBA}''},
\textsf{\doiref{10.1007/JHEP02(2011)025}{JHEP~1102,~025~(2011)}},
\texttt{\arxivref{1009.4118}{arxiv:1009.4118}}.

\bibitem{Beccaria:2009hg}
M.~Beccaria and G.~F.~De~Angelis,
\textit{``{On the wrapping correction to single magnon energy in twisted N=4
  SYM}''},
\textsf{\doiref{10.1142/S0217751X09047375}{Int.~J.~Mod.~Phys.~A24,~5803~(2009)%
}},
\texttt{\arxivref{0903.0778}{arxiv:0903.0778}}.

\bibitem{Bajnok:2009vm}
Z.~Bajnok, A.~Hegedus, R.~A.~Janik and T.~Lukowski,
\textit{``{Five loop Konishi from AdS/CFT}''},
\textsf{\doiref{10.1016/j.nuclphysb.2009.10.015}{Nucl.~Phys.~B827,~426~(2010)}%
},
\texttt{\arxivref{0906.4062}{arxiv:0906.4062}}.

\bibitem{Gunnesson:2009nn}
J.~Gunnesson,
\textit{``{Wrapping in maximally supersymmetric and marginally deformed N=4
  Yang-Mills}''},
\textsf{\doiref{10.1088/1126-6708/2009/04/130}{JHEP~0904,~130~(2009)}},
\texttt{\arxivref{0902.1427}{arxiv:0902.1427}}.

\bibitem{Ahn:2010yv}
C.~Ahn, Z.~Bajnok, D.~Bombardelli and R.~I.~Nepomechie,
\textit{``{Finite-size effect for four-loop Konishi of the beta-deformed N=4
  SYM}''},
\textsf{\doiref{10.1016/j.physletb.2010.08.056}{Phys.~Lett.~B693,~380~(2010)}},
\texttt{\arxivref{1006.2209}{arxiv:1006.2209}}.

\bibitem{Beccaria:2010kd}
M.~Beccaria, F.~Levkovich-Maslyuk and G.~Macorini,
\textit{``{On wrapping corrections to GKP-like operators}''},
\textsf{\doiref{10.1007/JHEP03(2011)001}{JHEP~1103,~001~(2011)}},
\texttt{\arxivref{1012.2054}{arxiv:1012.2054}}.

\bibitem{deLeeuw:2010ed}
M.~de~Leeuw and T.~Lukowski,
\textit{``{Twist operators in N=4 beta-deformed theory}''},
\texttt{\arxivref{1012.3725}{arxiv:1012.3725}}.

\bibitem{chapAmp}
R.~Roiban,
\textit{``Review of AdS/CFT Integrability, Chapter V.1: Scattering Amplitudes
  -- a Brief Introduction''},
\texttt{\arxivref{1012.4001}{arxiv:1012.4001}}.

\bibitem{Khoze:2005nd}
V.~V.~Khoze,
\textit{``{Amplitudes in the beta-deformed conformal Yang-Mills}''},
\textsf{\doiref{10.1088/1126-6708/2006/02/040}{JHEP~0602,~040~(2006)}},
\texttt{\arxivref{hep-th/0512194}{hep-th/0512194}}.

\bibitem{Oz:2007qr}
Y.~Oz, S.~Theisen and S.~Yankielowicz,
\textit{``{Gluon Scattering in Deformed N=4 SYM}''},
\textsf{\doiref{10.1016/j.physletb.2008.03.019}{Phys.~Lett.~B662,~297~(2008)}},
\texttt{\arxivref{0712.3491}{arxiv:0712.3491}}.

\bibitem{Mansson:2007jg}
T.~Mansson,
\textit{``{Is there a tower of charges to be discovered?}''},
\textsf{\doiref{10.1088/1751-8113/41/19/194014}{J.~Phys.~A41,~194014~(2008)}},
\texttt{\arxivref{0711.0931}{arxiv:0711.0931}}.

\bibitem{Bundzik:2005zg}
D.~Bundzik and T.~Mansson,
\textit{``{The general Leigh-Strassler deformation and integrability}''},
\textsf{\doiref{10.1088/1126-6708/2006/01/116}{JHEP~0601,~116~(2006)}},
\texttt{\arxivref{hep-th/0512093}{hep-th/0512093}}.

\bibitem{Bork:2007bj}
L.~V.~Bork, D.~I.~Kazakov, G.~S.~Vartanov and A.~V.~Zhiboedov,
\textit{``{Conformal Invariance in the Leigh-Strassler deformed N=4 SYM
  Theory}''},
\textsf{\doiref{10.1088/1126-6708/2008/04/003}{JHEP~0804,~003~(2008)}},
\texttt{\arxivref{0712.4132}{arxiv:0712.4132}}.

\bibitem{Mansson:2008xv}
T.~Mansson and K.~Zoubos,
\textit{``{Quantum Symmetries and Marginal Deformations}''},
\textsf{\doiref{10.1007/JHEP10(2010)043}{JHEP~1010,~043~(2010)}},
\texttt{\arxivref{0811.3755}{arxiv:0811.3755}}.

\bibitem{Mansson:2007sh}
T.~Mansson,
\textit{``{The Leigh-Strassler Deformation and the Quest for Integrability}''},
\textsf{\doiref{10.1088/1126-6708/2007/06/010}{JHEP~0706,~010~(2007)}},
\texttt{\arxivref{hep-th/0703150}{hep-th/0703150}}.

\bibitem{chapObserv}
C.~Kristjansen,
\textit{``Review of AdS/CFT Integrability, Chapter IV.1: Aspects of
  Non-planarity''},
\texttt{\arxivref{1012.3997}{arxiv:1012.3997}}.

\bibitem{chapABA}
M.~Staudacher,
\textit{``Review of AdS/CFT Integrability, Chapter III.1: Bethe Ans\"atze and
  the R-Matrix Formalism''},
\texttt{\arxivref{1012.3990}{arxiv:1012.3990}}.

\bibitem{Alday:2005ww}
L.~F.~Alday, G.~Arutyunov and S.~Frolov,
\textit{``{Green-Schwarz strings in TsT-transformed backgrounds}''},
\textsf{\doiref{10.1088/1126-6708/2006/06/018}{JHEP~0606,~018~(2006)}},
\texttt{\arxivref{hep-th/0512253}{hep-th/0512253}}.

\bibitem{Frolov:2005iq}
S.~A.~Frolov, R.~Roiban and A.~A.~Tseytlin,
\textit{``{Gauge-string duality for (non)supersymmetric deformations of N = 4
  super Yang-Mills theory}''},
\textsf{\doiref{10.1016/j.nuclphysb.2005.10.004}{Nucl.~Phys.~B731,~1~(2005)}},
\texttt{\arxivref{hep-th/0507021}{hep-th/0507021}}.

\bibitem{Prinsloo:2005dq}
A.~H.~Prinsloo,
\textit{``{gamma(i) deformed Lax pair for rotating strings in the fast motion
  limit}''},
\textsf{\doiref{10.1088/1126-6708/2006/01/050}{JHEP~0601,~050~(2006)}},
\texttt{\arxivref{hep-th/0510095}{hep-th/0510095}}.

\bibitem{Freyhult:2005ws}
L.~Freyhult, C.~Kristjansen and T.~Mansson,
\textit{``{Integrable spin chains with U(1)**3 symmetry and generalized
  Lunin-Maldacena backgrounds}''},
\textsf{\doiref{10.1088/1126-6708/2005/12/008}{JHEP~0512,~008~(2005)}},
\texttt{\arxivref{hep-th/0510221}{hep-th/0510221}}.

\bibitem{Bobev:2005ng}
N.~P.~Bobev, H.~Dimov and R.~C.~Rashkov,
\textit{``{Semiclassical strings, dipole deformations of N = 1 SYM and
  decoupling of KK modes}''},
\textsf{\doiref{10.1088/1126-6708/2006/02/064}{JHEP~0602,~064~(2006)}},
\texttt{\arxivref{hep-th/0511216}{hep-th/0511216}}.

\bibitem{McLoughlin:2006cg}
T.~McLoughlin and I.~Swanson,
\textit{``{Integrable twists in AdS/CFT}''},
\textsf{\doiref{10.1088/1126-6708/2006/08/084}{JHEP~0608,~084~(2006)}},
\texttt{\arxivref{hep-th/0605018}{hep-th/0605018}}.

\bibitem{Swanson:2007dh}
I.~Swanson,
\textit{``{A review of integrable deformations in AdS/CFT}''},
\textsf{\doiref{10.1142/S0217732307023614}{Mod.~Phys.~Lett.~A22,~915~(2007)}},
\texttt{\arxivref{0705.2844}{arxiv:0705.2844}}.

\bibitem{Beisert:2008tw}
N.~Beisert and P.~Koroteev,
\textit{``{Quantum Deformations of the One-Dimensional Hubbard Model}''},
\textsf{\doiref{10.1088/1751-8113/41/25/255204}{J.~Phys.~A41,~255204~(2008)}},
\texttt{\arxivref{0802.0777}{arxiv:0802.0777}}.

\bibitem{Kachru:1998ys}
S.~Kachru and E.~Silverstein,
\textit{``{4d conformal theories and strings on orbifolds}''},
\textsf{\doiref{10.1103/PhysRevLett.80.4855}{Phys.~Rev.~Lett.~80,~4855~(1998)}%
},
\texttt{\arxivref{hep-th/9802183}{hep-th/9802183}}.

\bibitem{Lawrence:1998ja}
A.~E.~Lawrence, N.~Nekrasov and C.~Vafa,
\textit{``{On conformal field theories in four dimensions}''},
\textsf{\doiref{10.1016/S0550-3213(98)00495-7}{Nucl.~Phys.~B533,~199~(1998)}},
\texttt{\arxivref{hep-th/9803015}{hep-th/9803015}}.

\bibitem{Ideguchi:2004wm}
K.~Ideguchi,
\textit{``{Semiclassical strings on AdS(5) x S**5/Z(M) and operators in
  orbifold field theories}''},
\textsf{\doiref{10.1088/1126-6708/2004/09/008}{JHEP~0409,~008~(2004)}},
\texttt{\arxivref{hep-th/0408014}{hep-th/0408014}}.

\bibitem{Beisert:2005he}
N.~Beisert and R.~Roiban,
\textit{``{The Bethe ansatz for Z(S) orbifolds of N = 4 super Yang- Mills
  theory}''},
\textsf{\doiref{10.1088/1126-6708/2005/11/037}{JHEP~0511,~037~(2005)}},
\texttt{\arxivref{hep-th/0510209}{hep-th/0510209}}.

\bibitem{Solovyov:2007pw}
A.~Solovyov,
\textit{``{Bethe Ansatz Equations for General Orbifolds of N=4 SYM}''},
\textsf{\doiref{10.1088/1126-6708/2008/04/013}{JHEP~0804,~013~(2008)}},
\texttt{\arxivref{0711.1697}{arxiv:0711.1697}}.

\bibitem{Sadri:2005gi}
D.~Sadri and M.~M.~Sheikh-Jabbari,
\textit{``{Integrable spin chains on the conformal moose}''},
\textsf{\doiref{10.1088/1126-6708/2006/03/024}{JHEP~0603,~024~(2006)}},
\texttt{\arxivref{hep-th/0510189}{hep-th/0510189}}.

\bibitem{Astolfi:2007uz}
D.~Astolfi, V.~Forini, G.~Grignani and G.~W.~Semenoff,
\textit{``{Gauge invariant finite size spectrum of the giant magnon}''},
\textsf{\doiref{10.1016/j.physletb.2007.06.002}{Phys.~Lett.~B651,~329~(2007)}},
\texttt{\arxivref{hep-th/0702043}{hep-th/0702043}}.

\bibitem{Arutyunov:2006gs}
G.~Arutyunov, S.~Frolov and M.~Zamaklar,
\textit{``{Finite-size effects from giant magnons}''},
\textsf{\doiref{10.1016/j.nuclphysb.2006.12.026}{Nucl.~Phys.~B778,~1~(2007)}},
\texttt{\arxivref{hep-th/0606126}{hep-th/0606126}}.

\bibitem{Ramadanovic:2008qd}
B.~Ramadanovic and G.~W.~Semenoff,
\textit{``{Finite Size Giant Magnon}''},
\textsf{\doiref{10.1103/PhysRevD.79.126006}{Phys.~Rev.~D79,~126006~(2009)}},
\texttt{\arxivref{0803.4028}{arxiv:0803.4028}}.

\bibitem{Bertolini:2002nr}
M.~Bertolini, J.~de~Boer, T.~Harmark, E.~Imeroni and N.~A.~Obers,
\textit{``{Gauge theory description of compactified pp-waves}''},
\textsf{\doiref{10.1088/1126-6708/2003/01/016}{JHEP~0301,~016~(2003)}},
\texttt{\arxivref{hep-th/0209201}{hep-th/0209201}}.

\bibitem{Astolfi:2008yw}
D.~Astolfi, G.~Grignani, T.~Harmark and M.~Orselli,
\textit{``{Finite-size corrections to the rotating string and the winding
  state}''},
\textsf{\doiref{10.1088/1126-6708/2008/08/099}{JHEP~0808,~099~(2008)}},
\texttt{\arxivref{0804.3301}{arxiv:0804.3301}}.

\bibitem{Mukhi:2002ck}
S.~Mukhi, M.~Rangamani and E.~P.~Verlinde,
\textit{``{Strings from quivers, membranes from moose}''},
\textsf{\doiref{10.1088/1126-6708/2002/05/023}{JHEP~0205,~023~(2002)}},
\texttt{\arxivref{hep-th/0204147}{hep-th/0204147}}.

\bibitem{Astolfi:2006is}
D.~Astolfi, V.~Forini, G.~Grignani and G.~W.~Semenoff,
\textit{``{Finite size corrections and integrability of N = 2 SYM and DLCQ
  strings on a pp-wave}''},
\textsf{\doiref{10.1088/1126-6708/2006/09/056}{JHEP~0609,~056~(2006)}},
\texttt{\arxivref{hep-th/0606193}{hep-th/0606193}}.

\bibitem{deRisi:2004bc}
G.~De~Risi, G.~Grignani, M.~Orselli and G.~W.~Semenoff,
\textit{``{DLCQ string spectrum from N = 2 SYM theory}''},
\textsf{\doiref{10.1088/1126-6708/2004/11/053}{JHEP~0411,~053~(2004)}},
\texttt{\arxivref{hep-th/0409315}{hep-th/0409315}}.

\bibitem{Gadde:2010zi}
A.~Gadde, E.~Pomoni and L.~Rastelli,
\textit{``{Spin Chains in N=2 Superconformal Theories: from the $Z_2$ Quiver to
  Superconformal QCD}''},
\texttt{\arxivref{1006.0015}{arxiv:1006.0015}}.

\bibitem{Gadde:2010ku}
A.~Gadde and L.~Rastelli,
\textit{``{Twisted Magnons}''},
\texttt{\arxivref{1012.2097}{arxiv:1012.2097}}.

\bibitem{Bershadsky:1998cb}
M.~Bershadsky and A.~Johansen,
\textit{``{Large N limit of orbifold field theories}''},
\textsf{\doiref{10.1016/S0550-3213(98)00526-4}{Nucl.~Phys.~B536,~141~(1998)}},
\texttt{\arxivref{hep-th/9803249}{hep-th/9803249}}.

\bibitem{Klebanov:1998hh}
I.~R.~Klebanov and E.~Witten,
\textit{``{Superconformal field theory on threebranes at a Calabi-Yau
  singularity}''},
\textsf{\doiref{10.1016/S0550-3213(98)00654-3}{Nucl.~Phys.~B536,~199~(1998)}},
\texttt{\arxivref{hep-th/9807080}{hep-th/9807080}}.

\bibitem{Schvellinger:2003vz}
M.~Schvellinger,
\textit{``{Spinning and rotating strings for N = 1 SYM theory and brane
  constructions}''},
\textsf{\doiref{10.1088/1126-6708/2004/02/066}{JHEP~0402,~066~(2004)}},
\texttt{\arxivref{hep-th/0309161}{hep-th/0309161}}.

\bibitem{Kim:2003vn}
N.~Kim,
\textit{``{Multi-spin strings on AdS(5) x T(1,1) and operators of N = 1
  superconformal theory}''},
\textsf{\doiref{10.1103/PhysRevD.69.126002}{Phys.~Rev.~D69,~126002~(2004)}},
\texttt{\arxivref{hep-th/0312113}{hep-th/0312113}}.

\bibitem{Wang:2005baa}
X.-J.~Wang,
\textit{``{Spinning strings on deformed AdS(5) x T(1,1) with NS B- fields}''},
\textsf{\doiref{10.1103/PhysRevD.72.086006}{Phys.~Rev.~D72,~086006~(2005)}},
\texttt{\arxivref{hep-th/0501029}{hep-th/0501029}}.

\bibitem{Benvenuti:2005cz}
S.~Benvenuti and M.~Kruczenski,
\textit{``{Semiclassical strings in Sasaki-Einstein manifolds and long
  operators in N = 1 gauge theories}''},
\textsf{\doiref{10.1088/1126-6708/2006/10/051}{JHEP~0610,~051~(2006)}},
\texttt{\arxivref{hep-th/0505046}{hep-th/0505046}}.

\bibitem{Benvenuti:2008bd}
S.~Benvenuti and E.~Tonni,
\textit{``{Giant magnons and spiky strings on the conifold}''},
\textsf{\doiref{10.1088/1126-6708/2009/02/041}{JHEP~0902,~041~(2009)}},
\texttt{\arxivref{0811.0145}{arxiv:0811.0145}}.

\bibitem{Giataganas:2009dr}
D.~Giataganas,
\textit{``{Semi-classical Strings in Sasaki-Einstein Manifolds}''},
\textsf{\doiref{10.1088/1126-6708/2009/10/087}{JHEP~0910,~087~(2009)}},
\texttt{\arxivref{0904.3125}{arxiv:0904.3125}}.

\bibitem{Dimov:2009ut}
H.~Dimov, M.~Michalcik and R.~C.~Rashkov,
\textit{``{Strings on the deformed $T^{1,1}$: giant magnon and single spike
  solutions}''},
\textsf{\doiref{10.1088/1126-6708/2009/10/019}{JHEP~0910,~019~(2009)}},
\texttt{\arxivref{0908.3065}{arxiv:0908.3065}}.

\bibitem{Giataganas:2010mj}
D.~Giataganas,
\textit{``{Semiclassical strings in marginally deformed toric AdS/CFT}''},
\texttt{\arxivref{1010.1502}{arxiv:1010.1502}}.

\bibitem{Benvenuti:2007qt}
S.~Benvenuti and E.~Tonni,
\textit{``{Near-flat space limit and Einstein manifolds}''},
\textsf{\doiref{10.1088/1126-6708/2008/02/022}{JHEP~0802,~022~(2008)}},
\texttt{\arxivref{0707.1676}{arxiv:0707.1676}}.

\bibitem{Maldacena:2006rv}
J.~M.~Maldacena and I.~Swanson,
\textit{``{Connecting giant magnons to the pp-wave: An interpolating limit of
  $AdS_5 \times S^5$}''},
\textsf{\doiref{10.1103/PhysRevD.76.026002}{Phys.~Rev.~D76,~026002~(2007)}},
\texttt{\arxivref{hep-th/0612079}{hep-th/0612079}}.

\bibitem{Sklyanin:1988yz}
E.~K.~Sklyanin,
\textit{``{Boundary Conditions for Integrable Quantum Systems}''},
\textsf{J.~Phys.~A21,~2375~(1988)}.

\bibitem{Arnaudon:2004sd}
D.~Arnaudon, J.~Avan, N.~Crampe, A.~Doikou, L.~Frappat and E.~Ragoucy,
\textit{``{General boundary conditions for the $sl(N)$ and $sl(M|N)$ open spin
  chains}''},
\textsf{J.~Stat.~Mech.~0408,~P005~(2004)},
\texttt{\arxivref{math-ph/0406021}{math-ph/0406021}}.

\bibitem{Beisert:2008cf}
N.~Beisert and F.~Loebbert,
\textit{``{Open Perturbatively Long-Range Integrable gl(N) Spin Chains}''},
\textsf{Adv.~Sci.~Lett.~2,~261~(2009)},
\texttt{\arxivref{0805.3260}{arxiv:0805.3260}}.

\bibitem{McGreevy:2000cw}
J.~McGreevy, L.~Susskind and N.~Toumbas,
\textit{``{Invasion of the giant gravitons from anti-de Sitter space}''},
\textsf{\doiref{10.1088/1126-6708/2000/06/008}{JHEP~0006,~008~(2000)}},
\texttt{\arxivref{hep-th/0003075}{hep-th/0003075}}.

\bibitem{Berenstein:2005vf}
D.~Berenstein and S.~E.~Vazquez,
\textit{``{Integrable open spin chains from giant gravitons}''},
\textsf{\doiref{10.1088/1126-6708/2005/06/059}{JHEP~0506,~059~(2005)}},
\texttt{\arxivref{hep-th/0501078}{hep-th/0501078}}.

\bibitem{Agarwal:2006gc}
A.~Agarwal,
\textit{``{Open spin chains in super Yang-Mills at higher loops: Some potential
  problems with integrability}''},
\textsf{\doiref{10.1088/1126-6708/2006/08/027}{JHEP~0608,~027~(2006)}},
\texttt{\arxivref{hep-th/0603067}{hep-th/0603067}}.

\bibitem{Hofman:2007xp}
D.~M.~Hofman and J.~M.~Maldacena,
\textit{``{Reflecting magnons}''},
\textsf{\doiref{10.1088/1126-6708/2007/11/063}{JHEP~0711,~063~(2007)}},
\texttt{\arxivref{0708.2272}{arxiv:0708.2272}}.

\bibitem{Mann:2006rh}
N.~Mann and S.~E.~Vazquez,
\textit{``{Classical open string integrability}''},
\textsf{\doiref{10.1088/1126-6708/2007/04/065}{JHEP~0704,~065~(2007)}},
\texttt{\arxivref{hep-th/0612038}{hep-th/0612038}}.

\bibitem{Berenstein:2005fa}
D.~Berenstein, D.~H.~Correa and S.~E.~Vazquez,
\textit{``{Quantizing open spin chains with variable length: An example from
  giant gravitons}''},
\textsf{\doiref{10.1103/PhysRevLett.95.191601}{Phys.~Rev.~Lett.~95,~191601~(20%
05)}},
\texttt{\arxivref{hep-th/0502172}{hep-th/0502172}}.

\bibitem{Berenstein:2006qk}
D.~Berenstein, D.~H.~Correa and S.~E.~Vazquez,
\textit{``{A study of open strings ending on giant gravitons, spin chains and
  integrability}''},
\textsf{\doiref{10.1088/1126-6708/2006/09/065}{JHEP~0609,~065~(2006)}},
\texttt{\arxivref{hep-th/0604123}{hep-th/0604123}}.

\bibitem{Ciavarella:2010tp}
A.~Ciavarella,
\textit{``{Giant magnons and non-maximal giant gravitons}''},
\textsf{\doiref{10.1007/JHEP01(2011)040}{JHEP~1101,~040~(2011)}},
\texttt{\arxivref{1011.1440}{arxiv:1011.1440}}.

\bibitem{Chen:2007ec}
H.-Y.~Chen and D.~H.~Correa,
\textit{``{Comments on the Boundary Scattering Phase}''},
\textsf{\doiref{10.1088/1126-6708/2008/02/028}{JHEP~0802,~028~(2008)}},
\texttt{\arxivref{0712.1361}{arxiv:0712.1361}}.

\bibitem{Murgan:2008fs}
R.~Murgan and R.~I.~Nepomechie,
\textit{``{Open-chain transfer matrices for AdS/CFT}''},
\textsf{\doiref{10.1088/1126-6708/2008/09/085}{JHEP~0809,~085~(2008)}},
\texttt{\arxivref{0808.2629}{arxiv:0808.2629}}.

\bibitem{Murgan:2008zu}
R.~Murgan and R.~I.~Nepomechie,
\textit{``{q-deformed $su(2|2)$ boundary S-matrices via the ZF algebra}''},
\textsf{\doiref{10.1088/1126-6708/2008/06/096}{JHEP~0806,~096~(2008)}},
\texttt{\arxivref{0805.3142}{arxiv:0805.3142}}.

\bibitem{Murgan:2009bx}
R.~Murgan,
\textit{``{A note on open-chain transfer matrices from q-deformed $su(2|2)$
  S-matrices}''},
\textsf{\doiref{10.1002/prop.200900080}{Fortschr.~Phys.~57,~895~(2009)}},
\texttt{\arxivref{0906.4361}{arxiv:0906.4361}}.

\bibitem{Ahn:2010xa}
C.~Ahn and R.~I.~Nepomechie,
\textit{``{Yangian symmetry and bound states in AdS/CFT boundary
  scattering}''},
\textsf{\doiref{10.1007/JHEP05(2010)016}{JHEP~1005,~016~(2010)}},
\texttt{\arxivref{1003.3361}{arxiv:1003.3361}}.

\bibitem{MacKay:2010ey}
N.~MacKay and V.~Regelskis,
\textit{``{Yangian symmetry of the Y=0 maximal giant graviton}''},
\textsf{\doiref{10.1007/JHEP12(2010)076}{JHEP~1012,~076~(2010)}},
\texttt{\arxivref{1010.3761}{arxiv:1010.3761}}.

\bibitem{Galleas:2009ye}
W.~Galleas,
\textit{``{The Bethe Ansatz Equations for Reflecting Magnons}''},
\textsf{\doiref{10.1016/j.nuclphysb.2009.04.024}{Nucl.~Phys.~B820,~664~(2009)}%
},
\texttt{\arxivref{0902.1681}{arxiv:0902.1681}}.

\bibitem{Okamura:2006zr}
K.~Okamura and K.~Yoshida,
\textit{``{Higher loop Bethe ansatz for open spin-chains in AdS/CFT}''},
\textsf{\doiref{10.1088/1126-6708/2006/09/081}{JHEP~0609,~081~(2006)}},
\texttt{\arxivref{hep-th/0604100}{hep-th/0604100}}.

\bibitem{Nepomechie:2009zi}
R.~I.~Nepomechie,
\textit{``{Bethe ansatz equations for open spin chains from giant
  gravitons}''},
\textsf{\doiref{10.1088/1126-6708/2009/05/100}{JHEP~0905,~100~(2009)}},
\texttt{\arxivref{0903.1646}{arxiv:0903.1646}}.

\bibitem{Bak:2008xq}
D.~Bak,
\textit{``{Zero Modes for the Boundary Giant Magnons}''},
\textsf{\doiref{10.1016/j.physletb.2009.01.035}{Phys.~Lett.~B672,~284~(2009)}},
\texttt{\arxivref{0812.2645}{arxiv:0812.2645}}.

\bibitem{Ahn:2007bq}
C.~Ahn, D.~Bak and S.-J.~Rey,
\textit{``{Reflecting Magnon Bound States}''},
\textsf{\doiref{10.1088/1126-6708/2008/04/050}{JHEP~0804,~050~(2008)}},
\texttt{\arxivref{0712.4144}{arxiv:0712.4144}}.

\bibitem{Palla:2008zc}
L.~Palla,
\textit{``{Issues on magnon reflection}''},
\textsf{\doiref{10.1016/j.nuclphysb.2008.09.021}{Nucl.~Phys.~B808,~205~(2009)}%
},
\texttt{\arxivref{0807.3646}{arxiv:0807.3646}}.

\bibitem{Ahn:2008df}
C.~Ahn and R.~I.~Nepomechie,
\textit{``{The Zamolodchikov-Faddeev algebra for open strings attached to giant
  gravitons}''},
\textsf{\doiref{10.1088/1126-6708/2008/05/059}{JHEP~0805,~059~(2008)}},
\texttt{\arxivref{0804.4036}{arxiv:0804.4036}}.

\bibitem{Correa:2009dm}
D.~H.~Correa and C.~A.~S.~Young,
\textit{``{Asymptotic Bethe equations for open boundaries in planar
  AdS/CFT}''},
\texttt{\arxivref{0912.0627}{arxiv:0912.0627}}.

\bibitem{Correa:2009mz}
D.~H.~Correa and C.~A.~S.~Young,
\textit{``{Finite size corrections for open strings/open chains in planar
  AdS/CFT}''},
\textsf{\doiref{10.1088/1126-6708/2009/08/097}{JHEP~0908,~097~(2009)}},
\texttt{\arxivref{0905.1700}{arxiv:0905.1700}}.

\bibitem{Bajnok:2010ui}
Z.~Bajnok and L.~Palla,
\textit{``{Boundary finite size corrections for multiparticle states and planar
  AdS/CFT}''},
\textsf{\doiref{10.1007/JHEP01(2011)011}{JHEP~1101,~011~(2011)}},
\texttt{\arxivref{1010.5617}{arxiv:1010.5617}}.

\bibitem{Ciavarella:2010je}
A.~Ciavarella and P.~Bowcock,
\textit{``{Boundary Giant Magnons and Giant Gravitons}''},
\textsf{\doiref{10.1007/JHEP09(2010)072}{JHEP~1009,~072~(2010)}},
\texttt{\arxivref{1007.1674}{arxiv:1007.1674}}.

\bibitem{Correa:2006yu}
D.~H.~Correa and G.~A.~Silva,
\textit{``{Dilatation operator and the super Yang-Mills duals of open strings
  on AdS giant gravitons}''},
\textsf{\doiref{10.1088/1126-6708/2006/11/059}{JHEP~0611,~059~(2006)}},
\texttt{\arxivref{hep-th/0608128}{hep-th/0608128}}.

\bibitem{Sadri:2003mx}
D.~Sadri and M.~M.~Sheikh-Jabbari,
\textit{``{Giant hedge-hogs: Spikes on giant gravitons}''},
\textsf{\doiref{10.1016/j.nuclphysb.2004.03.013}{Nucl.~Phys.~B687,~161~(2004)}%
},
\texttt{\arxivref{hep-th/0312155}{hep-th/0312155}}.

\bibitem{Hirano:2006ti}
S.~Hirano,
\textit{``{Fat magnon}''},
\textsf{\doiref{10.1088/1126-6708/2007/04/010}{JHEP~0704,~010~(2007)}},
\texttt{\arxivref{hep-th/0610027}{hep-th/0610027}}.

\bibitem{deMelloKoch:2009zm}
R.~de~Mello~Koch, T.~K.~Dey, N.~Ives and M.~Stephanou,
\textit{``{Hints of Integrability Beyond the Planar Limit}''},
\textsf{\doiref{10.1007/JHEP01(2010)014}{JHEP~1001,~014~(2010)}},
\texttt{\arxivref{0911.0967}{arxiv:0911.0967}}.

\bibitem{Drukker:2006xg}
N.~Drukker and S.~Kawamoto,
\textit{``{Small deformations of supersymmetric Wilson loops and open
  spin-chains}''},
\textsf{\doiref{10.1088/1126-6708/2006/07/024}{JHEP~0607,~024~(2006)}},
\texttt{\arxivref{hep-th/0604124}{hep-th/0604124}}.

\bibitem{Caputa:2010ep}
P.~Caputa, C.~Kristjansen and K.~Zoubos,
\textit{``{On the spectral problem of N=4 SYM with orthogonal or symplectic
  gauge group}''},
\textsf{\doiref{10.1007/JHEP10(2010)082}{JHEP~1010,~082~(2010)}},
\texttt{\arxivref{1005.2611}{arxiv:1005.2611}}.

\bibitem{Berenstein:2002zw}
D.~E.~Berenstein, E.~Gava, J.~M.~Maldacena, K.~S.~Narain and H.~S.~Nastase,
\textit{``{Open strings on plane waves and their Yang-Mills duals}''},
\texttt{\arxivref{hep-th/0203249}{hep-th/0203249}}.

\bibitem{Stefanski:2003qr}
B.~Stefanski,~Jr.,
\textit{``{Open spinning strings}''},
\textsf{\doiref{10.1088/1126-6708/2004/03/057}{JHEP~0403,~057~(2004)}},
\texttt{\arxivref{hep-th/0312091}{hep-th/0312091}}.

\bibitem{Chen:2004mu}
B.~Chen, X.-J.~Wang and Y.-S.~Wu,
\textit{``{Integrable open spin chain in super Yang-Mills and the plane-wave /
  SYM duality}''},
\textsf{\doiref{10.1088/1126-6708/2004/02/029}{JHEP~0402,~029~(2004)}},
\texttt{\arxivref{hep-th/0401016}{hep-th/0401016}}.

\bibitem{Chen:2004yf}
B.~Chen, X.-J.~Wang and Y.-S.~Wu,
\textit{``{Open spin chain and open spinning string}''},
\textsf{\doiref{10.1016/j.physletb.2004.04.013}{Phys.~Lett.~B591,~170~(2004)}},
\texttt{\arxivref{hep-th/0403004}{hep-th/0403004}}.

\bibitem{Erler:2005nr}
T.~Erler and N.~Mann,
\textit{``{Integrable open spin chains and the doubling trick in N = 2 SYM with
  fundamental matter}''},
\textsf{\doiref{10.1088/1126-6708/2006/01/131}{JHEP~0601,~131~(2006)}},
\texttt{\arxivref{hep-th/0508064}{hep-th/0508064}}.

\bibitem{Correa:2008av}
D.~H.~Correa and C.~A.~S.~Young,
\textit{``{Reflecting magnons from D7 and D5 branes}''},
\textsf{\doiref{10.1088/1751-8113/41/45/455401}{J.~Phys.~A41,~455401~(2008)}},
\texttt{\arxivref{0808.0452}{arxiv:0808.0452}}.

\bibitem{MacKay:2010zb}
N.~MacKay and V.~Regelskis,
\textit{``{On the reflection of magnon bound states}''},
\textsf{\doiref{10.1007/JHEP08(2010)055}{JHEP~1008,~055~(2010)}},
\texttt{\arxivref{1006.4102}{arxiv:1006.4102}}.

\bibitem{DeWolfe:2004zt}
O.~DeWolfe and N.~Mann,
\textit{``{Integrable open spin chains in defect conformal field theory}''},
\textsf{\doiref{10.1088/1126-6708/2004/04/035}{JHEP~0404,~035~(2004)}},
\texttt{\arxivref{hep-th/0401041}{hep-th/0401041}}.

\bibitem{McLoughlin:2005gj}
T.~McLoughlin and I.~Swanson,
\textit{``{Open string integrability and AdS/CFT}''},
\textsf{\doiref{10.1016/j.nuclphysb.2005.06.014}{Nucl.~Phys.~B723,~132~(2005)}%
},
\texttt{\arxivref{hep-th/0504203}{hep-th/0504203}}.

\bibitem{Susaki:2004tg}
Y.~Susaki, Y.~Takayama and K.~Yoshida,
\textit{``{Open semiclassical strings and long defect operators in AdS/dCFT
  correspondence}''},
\textsf{\doiref{10.1103/PhysRevD.71.126006}{Phys.~Rev.~D71,~126006~(2005)}},
\texttt{\arxivref{hep-th/0410139}{hep-th/0410139}}.

\bibitem{Susaki:2005qn}
Y.~Susaki, Y.~Takayama and K.~Yoshida,
\textit{``{Integrability and higher loops in AdS/dCFT correspondence}''},
\textsf{\doiref{10.1016/j.physletb.2005.07.058}{Phys.~Lett.~B624,~115~(2005)}},
\texttt{\arxivref{hep-th/0504209}{hep-th/0504209}}.

\bibitem{Okamura:2005cj}
K.~Okamura, Y.~Takayama and K.~Yoshida,
\textit{``{Open spinning strings and AdS/dCFT duality}''},
\textsf{\doiref{10.1088/1126-6708/2006/01/112}{JHEP~0601,~112~(2006)}},
\texttt{\arxivref{hep-th/0511139}{hep-th/0511139}}.

\end{thebibliography}
\bibliographystyle{nb}

\end{document}